\documentclass[sigconf]{acmart}

\usepackage{url}  %
\usepackage{graphicx} 
\usepackage{multirow}
\usepackage{multicol}
\usepackage{balance}

\robustify\bfseries

\usepackage{ccicons}
\usepackage{amsmath}

\setcopyright{iw3c2w3}

\copyrightyear{2019}
\acmYear{2019} 
\acmConference[WWW '19]{Proceedings of the 2019 World Wide Web Conference}{May 13--17, 2019}{San Francisco, CA, USA}
\acmBooktitle{Proceedings of the 2019 World Wide Web Conference (WWW '19), May 13--17, 2019, San Francisco, CA, USA}
\acmPrice{}
\acmDOI{10.1145/3308558.3313684}
\acmISBN{978-1-4503-6674-8/19/05}

\begin{document}
\title[Demographic Inference and Representative Population Estimates]{Demographic Inference and Representative Population Estimates from Multilingual Social Media Data}

\author{Zijian Wang}
\authornote{Work performed while at the University of Michigan}
\affiliation{\institution{Stanford University}}
\email{zijwang@stanford.edu}

\author{Scott A.\ Hale}
\affiliation{\institution{University of Oxford \&  \mbox{Alan Turing Institute}}}
\email{scott.hale@oii.ox.ac.uk}

\author{\mbox{David Adelani}}
\affiliation{\institution{MPI-SWS \&  \mbox{Saarland University}}}
\email{dadelani@mpi-sws.org}
\author{\mbox{Przemyslaw~A.~Grabowicz}}
\affiliation{\institution{MPI-SWS \& \mbox{University of Massachusetts}}}
\email{grabowicz@cs.umass.edu}

\author{Timo Hartmann}
\affiliation{\institution{GESIS Cologne}}
\email{timo.hartmann@gesis.org}

\author{Fabian Fl\"ock}
\affiliation{\institution{GESIS Cologne}}
\email{fabian.floeck@gesis.org}

\author{David Jurgens}
\authornote{Corresponding senior author \vspace{-1.5mm}}
\affiliation{\institution{University of Michigan}}
\email{jurgens@umich.edu}

\renewcommand{\shortauthors}{Z. Wang et al.}


\begin{abstract}

Social media provide access to behavioural data at an unprecedented scale and granularity. However, using these data to understand phenomena in a broader population is difficult due to their non-representativeness and the bias of statistical inference tools towards dominant languages and groups.
While demographic attribute inference could be used to mitigate such bias, current techniques are almost entirely monolingual and fail to work in a global environment.
We address these challenges by combining multilingual demographic inference with post-stratification to create a more representative population sample.  To learn demographic attributes, we create a new multimodal deep neural architecture for joint classification of age, gender, and organization-status of social media users that operates in 32 languages.  This method substantially outperforms current state of the art while also reducing algorithmic bias.  To correct for sampling biases, we propose fully interpretable multilevel regression methods that estimate inclusion probabilities from inferred joint population counts and ground-truth population counts. 
In a large experiment over multilingual heterogeneous European regions, we show that our demographic inference and bias correction together allow for more accurate estimates of populations  and make a significant step towards representative social sensing in downstream applications with multilingual social media.
\end{abstract}


\maketitle

\section{Introduction}

Data representing the attitudes and behaviors of a (national) base population is of great importance to policy making, social science research and commercial prediction tasks, but representative surveys are expensive and infrequent. Social media data has been proposed as a real-time and inexpensive way to measure social phenomena---this area of research is known as ``social sensing.'' Social media data has been used, with various levels of success, in areas such as infectious disease~\cite{Ginsberg2008Detecting,Lazer2014Parable,Generous2014Global}, migration and tourism~\cite{lamanna2018immigrant, Zagheni2017Leveraging,Barchiesi2015Quantifying}, and box office takings for films~\cite{Mestyan2013Early,deSilva2014Prediction}.

\begin{figure}[t]
    \centering
    \includegraphics[width=0.48\textwidth]{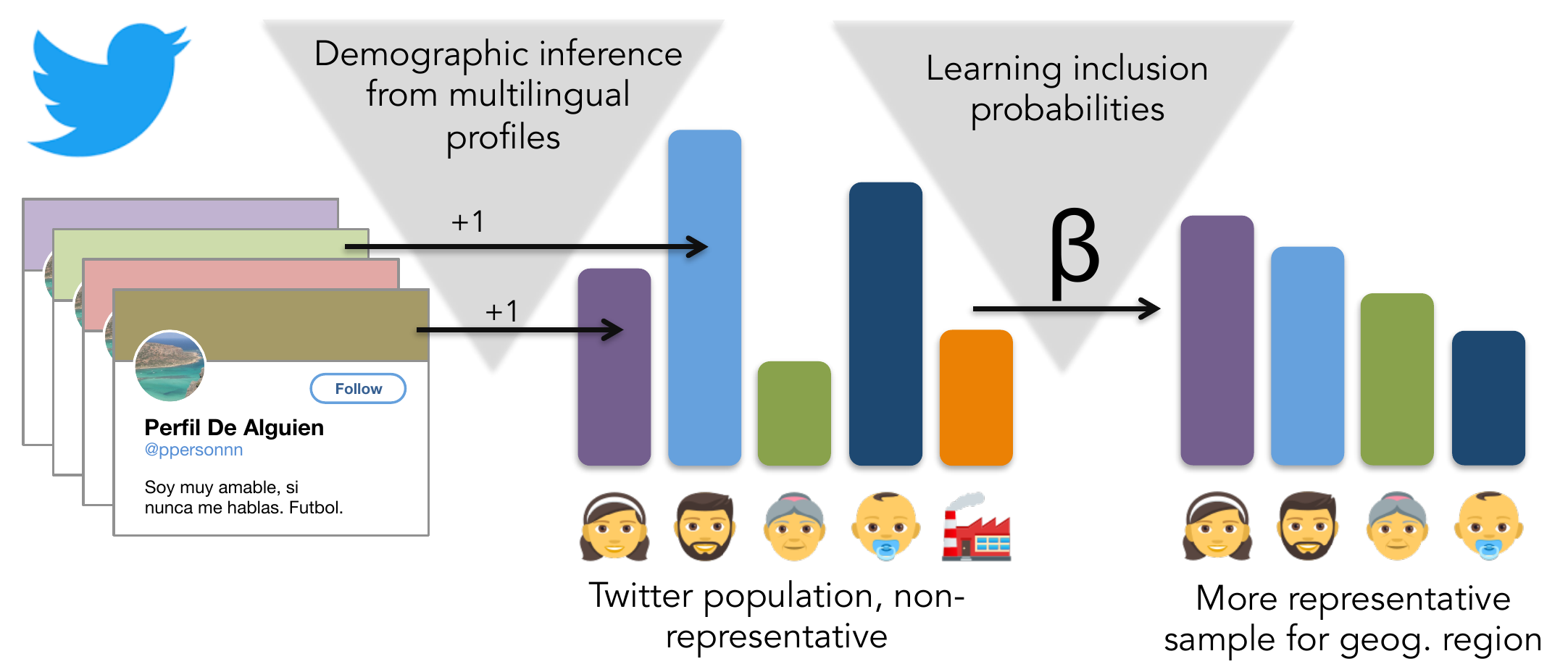}
    \caption{Our two-stage approach debiases non-representative Twitter data by (i) inferring demographics with state-of-the-art accuracy from multilingual, profile-only data and (ii) learning inclusion probabilities to create more representative samples for Europe-wide subregions}
    \label{fig:intro-figure}
    \vspace{-3mm}
\end{figure}
However, despite social media data being easily accessed, large in scale, and detailed, it is generally not representative of indicator variables measured in any broader
(offline) population due to various biases \cite{ruths2014social,jungherr2017normalizing}. Especially the self-selection of users joining and participating on any given social media platform is a large source of error. In the UK for instance, Twitter users are more likely to be male and young compared to the national population~\cite{Sloan2017Who}, and in the U.S. men and residents of densely populated areas are overrepresented \cite{hecht2014urban}, while there exists a mix of over- and undersampling on Twitter for users with specific racial backgrounds \cite{mislove2011understanding}. In general, however, the exact inclusion probabilities for a specific demographic group on any platform in a given country or region are unknown.
And while not all social sensing tasks require representative data~\cite{an2015whom}, recent studies have pointed out the fallacies in predicting phenomena via social sensing without controlling for sampling biases in social media data \cite{jungherr2012pirate,gayo2012wanted,gayo2013meta}, with research applications aiming  to draw inferences for nation-wide target populations often relying on such attribute data. Consequently, a systematic methodology for estimating them  is needed.

Survey analysis researchers have dealt with non-representative polls and non-response through sample re-weighting---with post-stratification  being a well-known technique \cite{bethlehem1987linear,holt1979poststratification}. These methods make use of basic demographics attributes: most prominently age, gender and location \cite{park2004bayesian,wang2015forecasting}.
Where demographic details are provided, post-stratification has been a valuable technique \cite{wang2015forecasting,Zagheni2017Leveraging}.
However, most social media platforms do not provide demographic data about their users, making it difficult to know or correct for such biases.
Although promising techniques for extracting demographic attributes from social media have been proposed, there exists no robust multilingual approach for this task. We address these challenges by inferring basic demographic attributes  and correcting for selection biases on a large sample of multilingual profile data from Twitter (Figure~\ref{fig:intro-figure}), one of the most used platforms for social sensing given its ease for obtaining data.

This work offers the following three contributions. First, we introduce a new  multilingual, multimodal, multi-attribute deep learning system for inferring demographics of users (\S\ref{sec:demo}). 
This pipeline is built from the ground up to enable inference in 32 different languages, addressing the need for such methods beyond English and outperforming state-of-the-art methods on the tasks of predicting the age, gender and is-organization state of Twitter users. %
We show this system reduces algorithmic bias with respect to skin tone over the best-performing commercial systems. %
Second, we formalize a statistical framework for models that debias non-representative samples (\S\ref{sec:debiasing}). Our models explicitly learn per-stratum inclusion probabilities from data, whereas typical post-stratification methods either assume their knowledge or focus on obtaining a post-stratified estimator of a response variable.
Third, in real-world evaluations, we show that our debiasing models significantly reduce the error rate in comparison to a model without post-stratification on the inferred demographic attributes (\S\ref{sec:twitter}). %
In these ways, we make a significant step towards \textit{representative social sensing} in downstream applications using multilingual social media.

\section{Demographic Inference}
\label{sec:demo}

We propose a new demographic inference model for three attributes: gender, age, and a binary organization indicator  (``is-organization'').  The first two attributes, gender and age, are widely reported in census data and are core features to accurately measure  demographic biases.   The third attribute was selected based on a known confounder for people-based studies of online platforms: the presence of accounts belonging to organizations \cite{alzahrani2018finding,mccorriston2015organizations}, which shall be distinguished from individual accounts. Next, we describe the model, its data, and its evaluation.

\subsection{Classification Task}

We consider the task of estimating population count in Twitter given a stream of user tweets.  This scenario is motivated by the common research use case for social sensing on Twitter, where a researcher consumes one of the Twitter streams (e.g., the 10\% or 1\% sample streams) and desires to make some analysis from the tweets of a broader phenomena that extends beyond Twitter.
In such a scenario, many users are expected to appear rarely--possibly only once--which hinders the use of existing demographic methods that rely upon large amounts of text to infer gender \cite{tannen1991you,tannen1993gender,kendall1997gender,coates1998language,eckert2003language,lakoff2004language,coates2015women,chen2015comparative,zhang2016your} and age \cite{rosenthal2011age,goswami2009stylometric,rao2010classifying,nguyen2011author,nguyen2013old,sap2014developing}.

Instead, we avoid making inferences based on tweets and focus on using only the information associated with a user's account, which allows our method to easily scale to large volumes of users without the need for significant quantities of text.
Not stratifying users based on tweets  has the added advantage of  preventing downstream demographic biases to any social sensing tasks that analyze the same language. %
Ultimately, our method uses four sources of information, username, screen name, biography, and profile image, and is designed to operate effectively on multilingual data.

We formalize the three classification tasks as follows.  Gender and is-organization are modeled as binary classification tasks.\footnote{We base our notion of gender as one of  performance \cite{defrancisco2013gender}, in which individuals adapt their style, name, and picture to (de)emphasize certain aspects of their gender identity \cite{eckert2008variation}.  While our training data uses self-declared binary gender identities, the M3 model reports gender identity as a continuum using prediction probability  along a graded scale.  We recognize that the current approach does not yet support non-binary gender identities, which we view as an important future task for full representation.
}  %
Age is known to be a difficult inference task in social media \cite{nguyen2013old,zhang2016your}.  Here, we opt to categorize age in four levels: $\le18$, $(18, 30)$, $[30, 40)$, $[40, 99)$.  Our choice is motivated by (i) our ability to align these age ranges with census data reports and  surveys \cite{harmonised2015ukstats, implementation2009usde}, and (ii) the difficulty of the task (even for humans) when finer-grained age classes are used \cite{dehon2001other,van2014age}, while (iii) still making the resulting classes amenable to downstream tasks.  
While other attributes are possible for use in debiasing (e.g., education, income, etc.), ground truth population statistics for these attributes are not widely available in censuses, making it difficult to train and evaluate any model.

\begin{figure}[tb]
    \centering
    \includegraphics[width=0.40\textwidth]{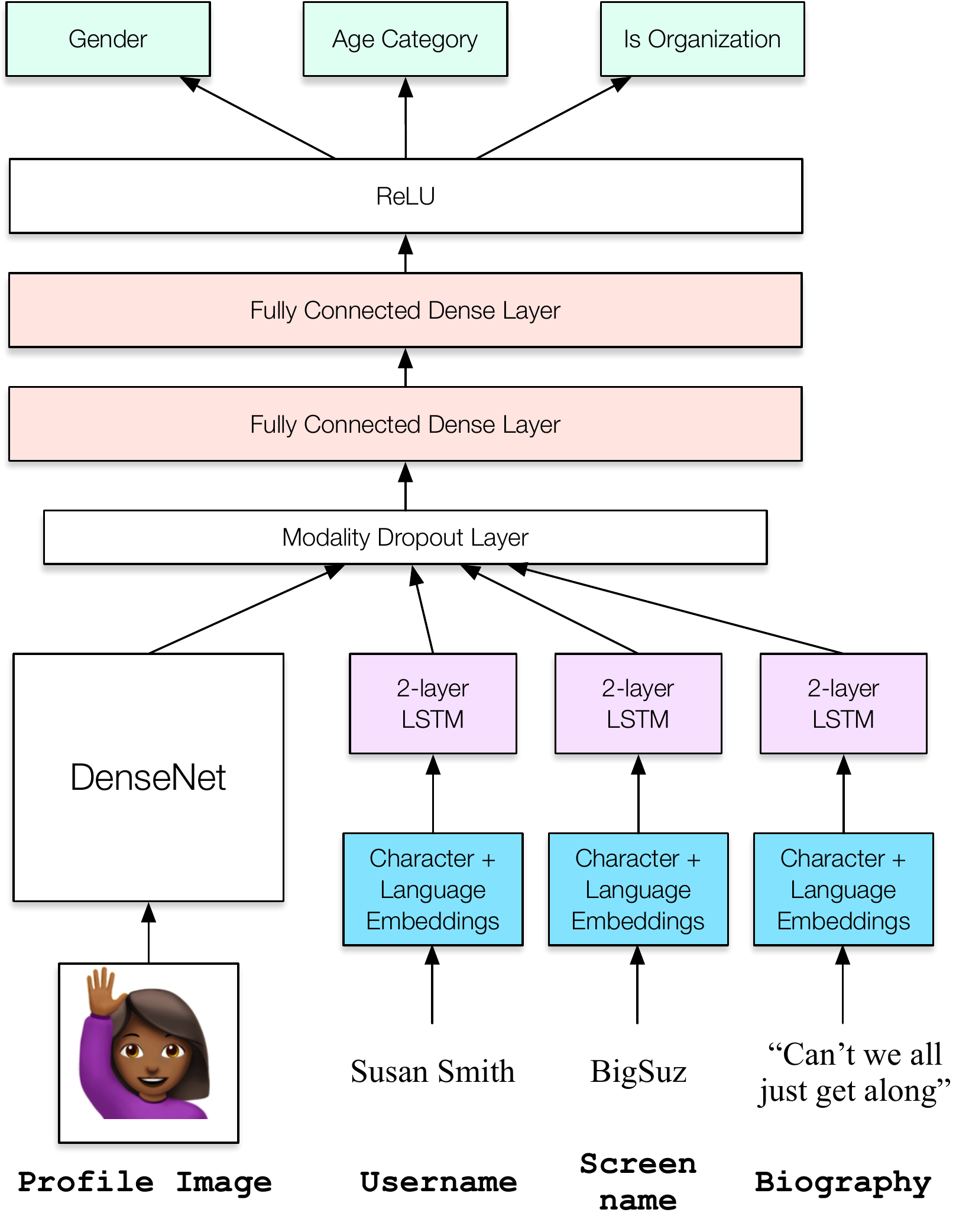}
    \caption{The \textbf{M3} model for inferring gender, age, and organization-identity from image and text data. }
    \label{fig:m3-model}
    \vspace{-7mm}
\end{figure}

\subsection{Method and Training Procedure}

Demographic attributes are often expressed in both image and text.  We propose a new multimodal inference model with a novel training technique that leverages both of these sources to significantly augment our predictive accuracy. We first describe the model and each modality and then discuss the training procedure.

\subsubsection{Model}

The full model comprises two separate pipelines for processing a profile image and each of the three text sources of information, as shown in Figure \ref{fig:m3-model}.  Here, we use a shared pipeline for all three attributes, using multi-task learning for the output.  We refer to this as the \textbf{M3} model, after its multimodal, multilingual, and multi-attribute abilities.  Following, we describe the architectures of each modality and how they are combined.

\noindent \emph{Image Model}
The image classifier was constructed using DenseNet\ \cite{huang2017densely} based on initial performance experiments over  current state-of-the-art vision models on our data. We scale all profile images to 224x224 %
to meet the input size of the model. 

\noindent \emph{Text Model}
Separate character-based neural models were trained for each text input and then a multi-input model was tuned based on the weights from each single model.  Embeddings for username and biography were limited to the  3000 most frequent characters in our training corpus, with remaining characters represented with an encoding for their unicode category; screen name are embedded in the ASCII range. We use 2-stack bidirectional  character-level Long-short Term Memory (LSTM) architecture \cite{hochreiter1997long} to capture necessary information.  To incorporate potential different meanings in different languages for the same character, we concatenate a separate trainable language embedding to each character embedding and add a fully connected layer before being fed into an LSTM. 
The fusion of character-based input with language embeddings provides two critical benefits that let us scale to the linguistic diversity seen in global platforms like twitter. (1) Character-based RNNs are able to capture morphological regularity within a language and substantially reduce the parameter space due to the reduced vocabulary size of the model (i.e., the set of characters) as compared with word-based models with orders-of-magnitude larger vocabularies \cite{kim2016character,faruqui2016morphological,chung2016character}.   (2) By fusing language embeddings with character embeddings, the joint embedding lets us represent shared linguistic regularity \textit{across} languages, which lets us better scale to the multilingual environment. 

\noindent \emph{Full Multimodal Model}
To construct the full model, separate models are fully trained for image and text for all attributes.  Then, the softmax output layer of each is removed and the two models are joined into a new model with a modality drop-out layer (described next), two fully connected layers of size 2048, a Rectified Linear Unit (ReLU) layer \cite{nair2010rectified}, and separate output layers for each task.  This full model is then fine-tuned using the pretrained single-modality models' parameters for initialization.

\subsubsection{Training Procedure}

Rather than training M3 end-to-end initially, we combine three  procedures for the training process: co-training, multilingual data augmentation, and modality drop out. %

\noindent \emph{Co-Training}
Both image and text can have clear signal of demographics.  Given the potentially-limited training data for speakers of less common languages, we take advantage of the multiple views of the same users by incorporating co-training: a semi-supervised learning technique. Co-training first learns a separate classifier for each view using any labeled examples. The most confident predictions of each classifier on the unlabeled data are then used to iteratively construct additional labeled training data \cite{blum1998combining}.  Ardehaly and Culotta \cite{ardehaly2017co} recently showed co-training was effective for demographic classification in social media data using  images and text in a graph-based co-training learning procedure.  %

In comparison to the variance of gender and age expressions in textual data across sociolinguistic contexts \cite{schler2006effects,bamman2014gender}, image data provides a more universal depiction, acting as an analog to a universal pivot language in machine translation \cite{wu2007pivot}.  Thus, we train the image-based portion of the M3 model first, which identifies high confidence labeled data from an unlabeled multilingual dataset.  Second, the text-based portion of the M3 model is trained using the original dataset plus the text data associated with any high-confidence image-based classifications.  Finally, we train the full M3 model with the training dataset and all high-confidence predictions from the unlabeled data.  All models are trained as multi-task, where a common input representation and architecture is used to  predict each output tasks. The probabilities for each task were represented using a top softmax layer. 

\noindent \emph{Multilingual Data Augmentation}
The M3 model is designed to operate in online environments on tens of languages.  However, few approaches and little labeled data exists for this task for languages other than English, with some exceptions \cite{ciot2013gender,goot2018bleaching,rangel2018overview}.  Even with co-training, the relative differences in number of speakers between languages (e.g., English vs.\ Slovenian) create a substantial class imbalance which would prevent the model from effectively learning signals of each attribute in users' biographies written in less-frequent languages.  Therefore, we perform data augmentation by using automatic machine translation of the training data.  

Given the scale of our training data (described later in \S\ref{sec:m3-data}), the use of online translation services is prohibitive both in terms of time and cost.  Instead, we opt to use word-based translation using the MUSE bilingual dictionaries \cite{conneau2017word}.  For each instance in the training data, all words in the bilingual dictionary are replaced with the translation.   A translated instance is only kept if 80\% or more of its words could be translated.  While word substitution typically produces translations that are more incorrect and less fluent, we observed that the remaining biography translations were often of reasonable quality due to the declarative nature of biographies, which often consist primarily of lists of readily-translatable noun descriptions or attributes of the user without using full grammatical sentences.  Such translations also largely capture topical and semantic regularity, even if the morphology is not precise.

\noindent \emph{Modality Drop-out}
The co-training model uses a conjoining architecture with pre-trained weights from the image model and the multi-input text model.  To facilitate the model using each source of information, we introduce a new technique, modality dropout, that can fully hide a source of information (e.g., the biography) during training. 
We add an extra input dropout with a probability of 0.25 (i.e., keep 3 out of 4 input sources on average) to better regularize the model and support use cases where not all 4 input sources are available. When doing dropout, images were replaced by a random matrix while text was turned to a specific empty embedding. %
\vspace{-3mm}
\subsection{Data}
\label{sec:m3-data} 

Training and evaluating M3 uses five distinct datasets: (1) a large  dataset of Twitter profiles whose gender and age are heuristically identified, predominantly in English, (2) a curated dataset of Twitter accounts belonging to organizations, (3) an image dataset of faces from IMDB and Wikipedia, (4) a massive unlabeled set of users, and (5) a crowdsourced dataset for all three attributes that spans all 32 languages in our study.  We describe each next.

\subsubsection{Heuristically-identified  users}
Heuristically-identified  users were drawn from a 10\% sample of Twitter from 2014 to 2017 on Twitter, where roughly 40\% of the users are English speakers \cite{hale2014global}. Here, we identify gender and age from fixed expression in users' biographies that signal age or gender, e.g., ``mother of two wonderful kids'' or ``26 y/o dude'' using a gender- or age-signaling word.
Age data was further augmented by identifying tweets wishing another user a happy birthday (cf. \citet{al2012homophily}).  These pattern-based approaches are known to be highly precise for certain attributes when properly constructed \cite{beller2014ma,bergsma2013using}.  Patterns were created for five languages: English, Spanish, French, German, and Swedish.\footnote{Languages were selected to capture linguistic diversity and where high precision demographic-identify patterns  could be created easily.}
Organization accounts were manually curated by identifying 676 Twitter lists that primarily contained organizations, such as non-profits, local clubs, companies, or municipal services.  These lists resulted in 59.92K unique organization accounts.  
For heuristically-identified users and organization accounts, we collect their current profile image, screen name, username, and biography.  User age is adjusted to the present day.  For heuristically-identified users, the biography is altered to replace the gender- or age-indicating word with a special token so that the model is forced to look for additional cues to recognize each attribute. 
The IMDB-WIKI dataset \cite{rothe2016deep} consists of 523,051 images from headshots of actors in the IMDB and profile pictures in Wikipedia.  This dataset provides an auxiliary source of information for fine-tuning the image-based part of the M3 model.  

These three datasets constitute our \textit{initial data} prior to performing the three-step annotation procedure.   Table~\ref{tab:data} shows the detailed statistics of the data for each step.

Co-training depends on having access to a large set of unlabeled users whose high-confidence labels from one modality (e.g., profile image) can be used to augment the training set for another modality.  Here, we collect 36.97M profiles speaking one of the 32 languages commonly spoken in Europe but for which we have no groud truth label.  Languages were identified using CLD2 \cite{cld2}.  Note that users in this unlabeled dataset will only become training instances if one view's classifier (text or image) labels them with high confidence.

\begin{table}[tb]

    \centering
     \caption{Sizes of datasets used for training and testing.  Incremental numbers are shown for the outputs of data-augmentation from co-training and translation.}

    \begin{tabular}{rrrr} 
         Dataset & {\textbf{Gender}} & \textbf{Age } & \textbf{Is-Org.} \\
         \hline
         \textit{Initial} & 3.98M & 1.20M &59.92K\\
         After Image Co-Training & +2.26M & +0.62M & +6.31M\\
         After Translation Augmentation %
            & +5.74M & +1.54M & +4.74M \\
         After Text Co-Training & +8.28M & +0.78M & +11.87M\\

         \textit{Final}&  14.53M & 2.61M & 23.86M\\
         Held-out Evaluation & 0.38M & 0.36M & 6.91K\\
    \end{tabular}

    \vspace{-0.4cm}
    \label{tab:data}
\end{table}

\subsubsection{Crowdsourcing Data}

Current gender, age, and organization datasets are nearly all produced for English-speaking users, which prevents us from evaluating the M3 model in the multilingual environment.  Therefore, we constructed a new dataset of up to 200 randomly-selected Twitter users speaking one of each of the 32 languages in our study. %
Three annotators on Figure Eight\footnote{\url{https://www.figure-eight.com/}} %
were shown the usernames, screen names, short biographies, and profile images of each Twitter account. %
We instructed workers to determine gender and age of a given user from two drop-down menus based only on this information. The selection options regarding gender also included a category for profiles belonging to multiple persons, and to a non-person (e.g., organization/bot). %
Regarding age, annotators were given seven categories to choose from: $<=18$, $(18, 30)$, $[30, 40)$, $[40, 50)$, $[50, 60)$, $[60, 70)$, $>=69$.  They could select ``don't know'' for all questions, if they felt not enough information was available.
We employed Figure Eight's selection mechanism for speakers of the target language of the set to be annotated, or chose workers by their country of residence, if one of that country's main spoken languages was identical with the target language of the task.
Additionally, each job included a set of 30 test questions, 
annotated as a gold standard by the authors. These were used to screen for spam but also for any non-speakers of the target language remaining after our first filter; to this end, at least 10 questions were only solvable by sufficiently understanding the target language. %

Annotation reliability was measured for each Figure Eight job using Krippendorff's $\alpha$ \cite{krippendorff2011computing}. %
For the computations of $\alpha$, we excluded all English test-questions from the 31 non-English jobs, profiles with only one annotation, and included ``don't know'' answers as an answer option.
To establish a measure of how well workers did on the age  classification in respect to the age brackets classified by M3, we collapsed age bracket encodings over 40 into one category. %
Across all languages, the mean %
$\alpha$ for the gender annotations lies at  0.81 %
and at 0.57 regarding the age brackets, hinting at the high difficulty of the Twitter profile age classification task for humans. 
For a separate measure, we recoded all gender and organization annotations into the two categories ``is organization'' and ``is not organization''
and calculated Krippendorff's $\alpha$ based on these data, resulting in a mean of approximately 0.75 across all languages, showing that workers could identify non-personal accounts reasonably well. %
The final dataset is constructed from all instances where at least two annotators agreed on the label for a particular attribute (majority vote), resulting in 4,732 profiles across 32 languages.

\subsubsection{Data Partitions}
 
All self-reported labeled data is partitioned into 80\% train, 10\% development, and 10\% test splits.   No official train/test/dev splits are available for the IMBD-WIKI datasets so we perform our own partition with the same percentages as for self-reported.  During model training, only the development set for self-reporting users are used for model selection.
Owing to its small size, crowdsourcing data was only used for testing purposes.
For the organizational data, we create a full dataset with non-organizational accounts by randomly sampling accounts from the heuristically labeled data to attain a 1:9 ratio, following the report in McCorriston et al. \cite{mccorriston2015organizations} that organizations make up 10\% of the accounts on Twitter.

\subsubsection{Co-Training Setup}

The training data is balanced for each stage, where we oversampled gender and organization status, and undersampled the two most-frequent classes (0-18 and 19-29) while oversampling the other two age classes.
High confidence thresholds were set to 0.9 for gender and organization status in both the image and text modalities. As a four-class classification task, the high-confidence for age is set to 0.7 in the image classifier, based on early examination of the prediction quality.  We observed that age is more readily predictable in text and therefore used a threshold of 0.9 to keep quality high while also providing  a sufficient number of new accounts for co-training.
 
Separate image and text models are each trained for 10 epochs.
Due to the imbalanced number of instances per attribute (Table~\ref{tab:data}), each epoch contains 10000 steps where each step contains a mini-batch of 128 instances balanced across the three attributes.  This setup ensures the model sees instances of the smaller datasets (age and organization).  The image model is trained first, and its high-confidence instances are then used to train the text model.   The final co-training model was trained for 5 epochs with input dropout and another 3 epochs without input dropout for better convergence. Amsgrad optimizers \cite{reddi2018convergence} were used in all training process, with a learning rate of 0.001 for ground-up models and 0.0005 for fine-tuning models. Text models were trained on one GPU, while the image and co-training models were trained in parallel on 4 GPUs.  
The full pipeline was built in PyTorch \cite{paszke2017automatic}.

\subsection{Evaluations}
 
Accurate inference of representative attitudes from social media relies upon accurate stratification approaches.  Here, we test the M3 model against current state of the art systems to describe the potential for error to affect demographic bias.

\subsubsection{Comparison Systems and Data}

The M3 model is compared with state-of-the-art systems from each modality and attribute.  For images, we compare with Face++ \cite{jung2017inferring} and Microsoft Face API \cite{microsoft18face} on age and gender performance.  For text, we compare with three current state-of-art systems for inferring gender from usernames, genderperformr \cite{wang2018its}, demographer \cite{knowles2016demographer} and Jaech et al.~\cite{jaech2015your}, which could feasibly operate on names from multiple languages.
 
No multilingual organization recognition system exists; so, we limit our evaluation to the only publicly-available dataset of person-vs-organization which was scored by the Humanizr~\cite{mccorriston2015organizations} and Demographer~\cite{wood2018johns} systems.  This data consists of a uniform sample across Twitter accounts, which were approximately 10\% organizations.  Here, we recollect the user profiles for their 20,273 accounts, of which 18,587 (91.6\%) were still available as of October 2018, which matched the distribution originally reported.  Their model was evaluated on the dataset using cross-validation, whereas we treat the entire data as a test set and report performance. 

\subsubsection{Gender Recognition}

\begin{figure}[t]
    \centering
    \includegraphics[width=0.45\textwidth]{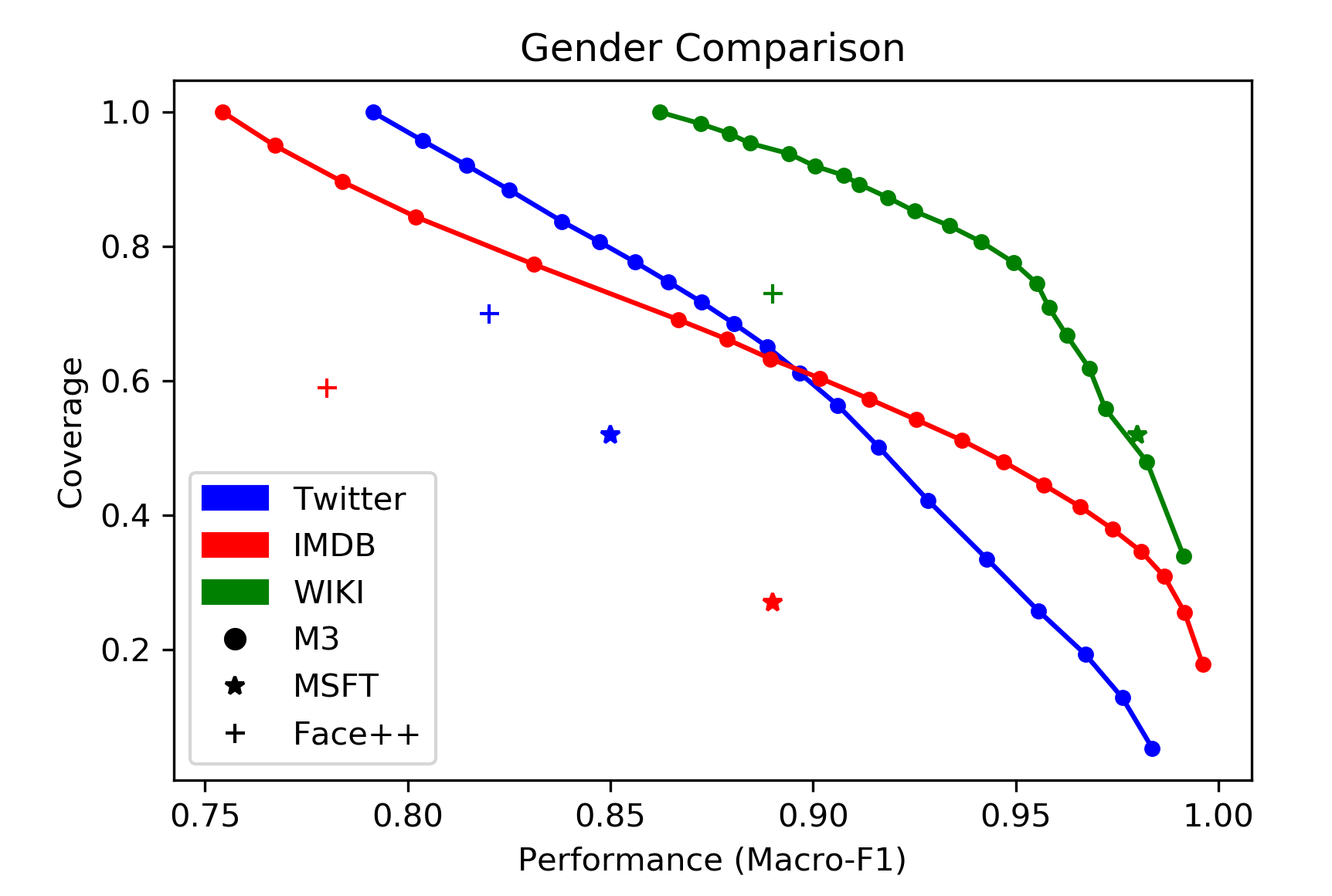}
    \vspace{-1mm}
    \caption{Model performances on gender classification for the three image datasets, as measured through F1. Coverage equals \% of returned results at varying confidence levels for M3 and is the maximum available for the commercial APIs. }
    \label{fig:genderface}
\end{figure}

M3 produces state-of-the-art gender recognition for each of the three image-based datasets (IMDB, WIKI, and Twitter), as seen in Figure~\ref{fig:genderface}.   M3  provides significant improvements in performance (F1) at each comparison system's level of recall.\footnote{The coverage reported is the maximum available from the commercial APIs, which did not find a face or did not provide a gender/age estimate for each photo. }  The one exception to this trend is the performance of the MSFT classifier on the WIKI dataset, which has highly similar performance to M3.  Given the MSFT's model large jump in performance, we hypothesize that their model may have been trained on Wikipedia data, although this cannot be confirmed.  We observed that the increase in coverage of M3 is due in part to the model learning non-facial attributes associated with more than one gender, e.g., certain sports with the male gender, which allow M3 to process profile photos that other models cannot.
 
On the text-based data, M3 outperforms all systems except one when using just the username and masking all other information, shown in Table \ref{tab:gender-text-perf}.   When the model is allowed to see all other text information, M3 performance improves substantially, indicating it can effectively fuse several sources of information.  While this latter setup uses more information than comparison systems, in practice, the Twitter API includes all text information used by M3 by default; so the only additional step is downloading the profile image.

\subsubsection{Age Recognition}

\begin{figure}[t]
    \centering
    \includegraphics[width=0.38\textwidth]{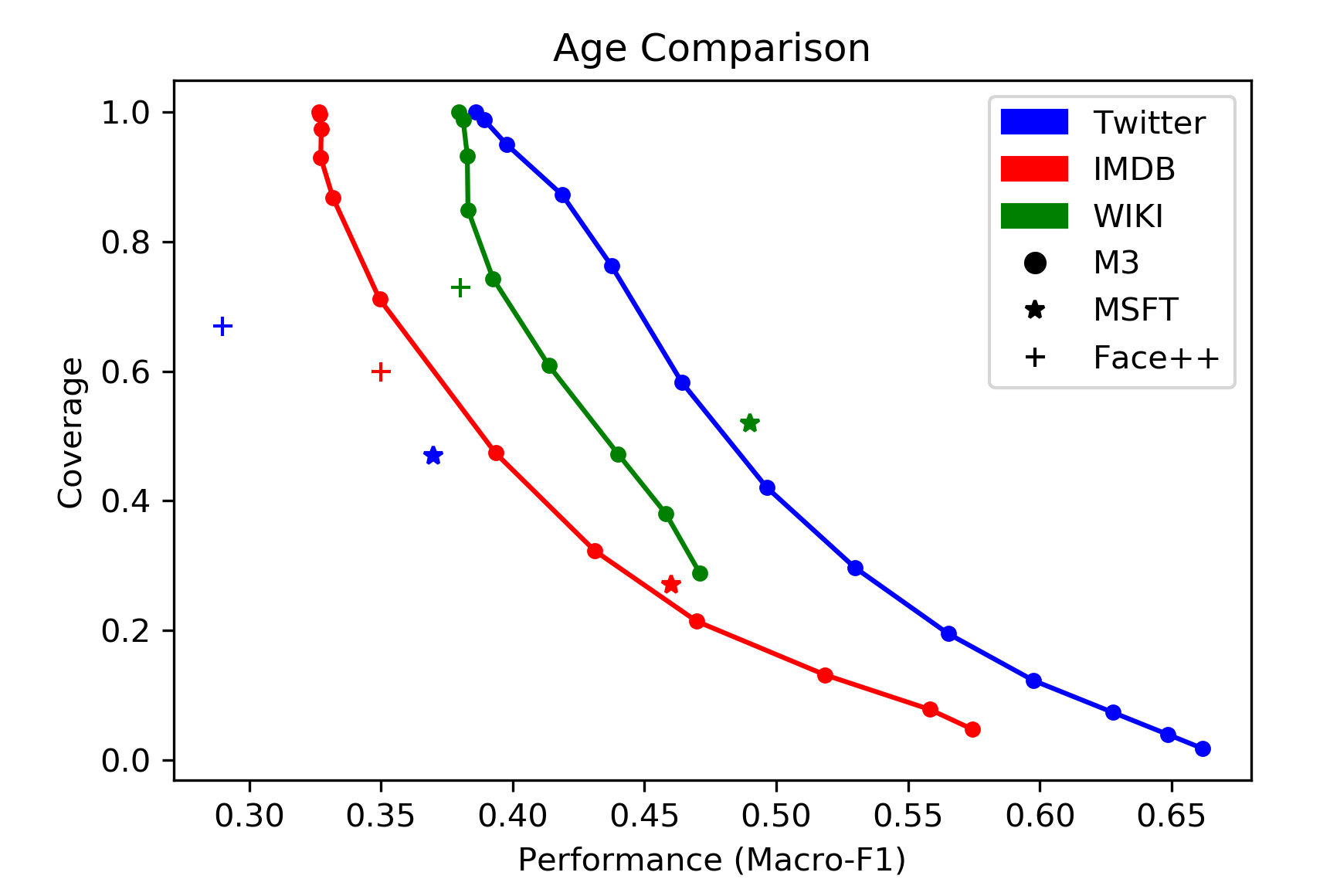}
     \vspace{-1mm}
    \caption{Model performances on age classification for three datasets, as measured through F1 on classified images and Coverage on the \% of images classified.}
    \label{fig:ageface}
\end{figure}

Age recognition is a significantly more difficult task in social media, as seen by the performances in Figure \ref{fig:ageface}.  The M3 model offers similar performance to commercial models on the IMDB and WIKI datasets, which primarily feature face-forward headshots in good lighting.  However, in real-world Twitter profiles (shown in blue), the M3 model substantially outperforms both in F1 at each of their respective coverage levels, with 0.16 and 0.11 absolute improvements over Face++ and Microsoft, respectively.

\begin{table}[b!]
 \caption{Performance comparison on gender using text methods. $^\dagger$Demographer only has a coverage of 0.974. }
    \begin{tabular}{rr@{}l}
         Method & {Macro-F1}\\
        \hline      
        M3: Full Text & \bfseries 0.907 \\
        M3: Username Only & 0.828\\
        GenderPerformr \cite{wang2018its} & 0.835 \\
        Demographer \cite{knowles2016demographer} & 0.781&$^\dagger$ \\
        \citet{jaech2015your} & 0.763 \\
    \end{tabular}
    \vspace{-5mm}
   
    \label{tab:gender-text-perf}
\end{table}

\subsubsection{Organization Recognition}

For any task that draws inferences about human actors, non-human and specifically organizational accounts are a major source of error. %
The results of Table~\ref{tab:org-perf} show that our model is able to identify these accounts with 16.3\% higher accuracy than the next closest system with no drop in performance at recognizing humans.  Results on our test set from our organization data (\S\ref{sec:m3-data}) in Table \ref{tab:m3-ablation} (top right) show similar performance, with an overall F1 of 0.898 that indicates high performance for both classes.

 \begin{table}[b!]
  \caption{Performance at recognizing organizational accounts, measured as accuracy per class, following \cite{mccorriston2015organizations}}
    \centering
    \begin{tabular}{rcc}
         Method & Person & Organization \\
        \hline         
        M3       & \textbf{0.986} & \textbf{0.807} \\
        Demographer \cite{wood2018johns} &  0.973 &  0.644 \\
        Humanizr \cite{mccorriston2015organizations} & 0.982 & 0.586 \\
    \end{tabular}
   
    \label{tab:org-perf}
      \vspace{-5mm}
\end{table}

\subsubsection{Multilingual Performance}
 
\begin{figure}
    \centering
    \vspace{-5mm}
    \includegraphics[width=.43\textwidth]{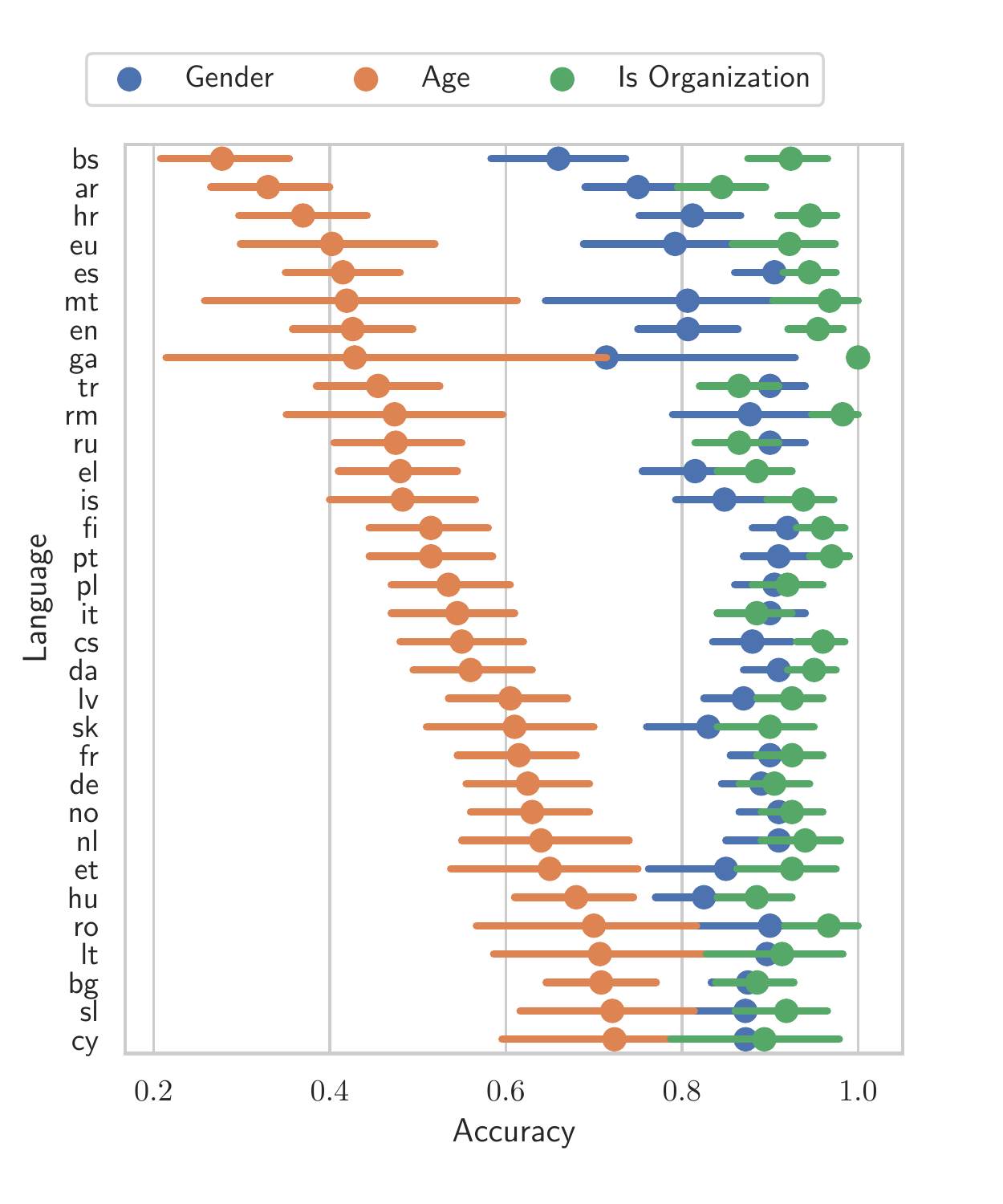}
    \vspace{-7mm}
    \caption{M3 performance on the multilingual dataset in 32 languages.  Bars show 95\% confidence intervals of  accuracy.}
    \label{fig:m3-multilingual}
\end{figure}

As a full test of M3, we evaluate it against our crowdsourced  account labels in 32 languages, shown in Table~\ref{tab:m3-ablation} top right.   Performance in this multilingual setting is on par with the performance on the primarily-English heuristically-labeled data, though age classification remains the most difficult task.  Figure \ref{fig:m3-multilingual} shows the performance per language for each attribute.  Performance at predicting gender and is-organization are similar for most languages, while age performance varies significantly by language from 0.28 F1 for Bosnian to 0.73 for Slovenian and Welsh.  %
These results together indicate that M3 is sufficiently accurate in the multilingual European environment.

\subsubsection{Ablation Study}

To examine which parts of the system are contributing to performance, we perform an ablation study by (i) restricting the model to one modality and (ii) training a model without using co-training or translation for data augmentation.  We test each model on (i) the heuristically-labeled and organisation data and (ii) the multilingual data, which is representative of the performance on the data used in our downstream task. 

The results shown in Table \ref{tab:m3-ablation} reveal two main trends.  First, the  model benefits from both modalities. The removal of textual information causes the biggest performance drop.  Second, although co-training and translation hurt performance in the heuristically-labeled data, they produced a substantially better model when evaluated on multilingual data.  Since the heuristically-labeled data is primarily in English,  translation potentially adds noise and forces the model to represent a larger space of inputs that are not represented in the test data.  This performance difference indicates that the two data augmentation techniques are highly beneficial when bootstrapping a model from mostly monolingual data.
 
 \begin{table}[b!]
 
    \centering
    \caption{Performance (Macro-F1) for (1) full M3 model, (2-3) model with modality dropouts, (4) model trained without  co-training or translation, (5-6) Random and and Majority-class baselines.  }
    \resizebox{0.49\textwidth}{!}{%
    \begin{tabular}{r ccc ccc}
        & \multicolumn{3}{c}{Heuristically-labeled} & \multicolumn{3}{c}{Multilingual} \\
         M3 Model & Gender & Age & Org-Status & Gender & Age & Org-Status \\
        \hline      
        Full & \textbf{0.918} & \textbf{0.522} & 0.898 & \textbf{0.915} & \textbf{0.425} & \textbf{0.898}\\
        w/ Text Dropout &  0.743 & 0.349 & 0.837 & 0.837 & 0.359& 0.892\\
        w/ Image Dropout & 0.905 & 0.493 & 0.823 & 0.862 & 0.372 & 0.822 \\
        w/o Co-T. \& Trans. &
        \textbf{0.918} & 0.449 & \textbf{0.917} &  0.875 & 0.325 &  0.660 \\
        \textit{Random} & 0.494 & 0.181 & 0.500 & 0.494 & 0.211 & 0.466\\
        \textit{Majority} & 0.377 & 0.214 & 0.500 & 0.334 & 0.187 & 0.210
    \end{tabular}
    }
    
    \label{tab:m3-ablation}
   \vspace{-5mm}
\end{table}

\subsubsection{Test for Algorithmic Bias}
 
Image-based models for gender recognition are known to suffer from algorithmic bias in recognizing darker-skinned individuals \cite{buolamwini2018gender}.
This bias is thought to be a result of non-representative training data the underreprentation of darker skin tones in the training data.  As our focus is on European countries, such a bias could be present in the M3 data.  However, because our co-training procedure uses a large unlabeled set of users from across the globe, these users can potentially provide a more representative sample and reduce algorithmic disparity.  To test for algorithmic bias, we compare the M3 model's performance on the Gender Shades dataset \cite{buolamwini2018gender}, which contains 3964 gender-annotated facial images balanced across light and dark skin tones.  In this earlier algorithmic audit, the Microsoft (MSFT) and Face++ gender classifiers performed substantially worse on women, especially darker skinned women.  

The gender inference performance, shown in Table~\ref{tab:gender-shades}, reveals that M3 has substantially less algorithmic bias relative to the two best-performing  commercial systems also tested on Gender Shades.  Our model significantly improves performance on dark-skinned women in comparison to the other two systems which had  the largest performance disparity  on that demographic.   However, M3 is least accurate on darker-skinned males, which indicates that additional work is needed to reach performance parity in skin tones.
Nevertheless, this analysis also indicates that M3 is more suitable than existing systems for operating in the global environment and increases the robustness of downstream social sensing applications.

\begin{table}[t!]
    \caption{Performance on the Gender Shades dataset (cf. Table 4 in \cite{buolamwini2018gender}).  Gender classification performance as measured by the positive predictive value (PPV), true positive rate (TPR), and false positive rate (FPR). }
    \resizebox{0.49\textwidth}{!}{%

\begin{tabular}{r c ccc cc cccc }    
    Classifier & Metric & {All} & {F} & {M} & {Darker} & {Lighter} & {DF} & {DM} & {LF} & {LM} \\
    \hline
\multirow{3}{*}{MSFT} & PPV & 93.7 & 89.3 & 97.4 & 87.1 & 99.3 & 79.2 & 94.0 & 98.3 &  100.0 \\
 & TPR & 93.7 &96.5 &91.7 &87.1 &99.3 &92.1 &83.7 & 100.0 & 98.7\\
 &  FPR & 6.3 &8.3 &3.5 &12.9 &0.7 &16.3 &7.9 &1.3&  0.0\\
\hline
\multirow{3}{*}{Face++} & PPV & 90.0 & 78.7 & 99.3 & 83.5 & 95.3 & 65.5 &  99.3 & 94.0 & 99.2 \\
 &TPR& 90.0 &98.9 &85.1 &83.5 &95.3 &98.8 &76.6 &98.9 &92.9 \\
 & FPR &  10.0& 14.9 &1.1 &16.5& 4.7 &23.4 & 1.2 & 7.1 &1.1\\

\hline

\multirow{3}{*}{M3}
 & PPV & 96.7 & 99.4 & 94.5 & 94.0 & 99.3 & 99.6 & 89.3 & 99.3 & 99.2 \\ 
 & TPR & 96.5 & 92.8 & 99.6 & 93.4 & 99.3 & 86.0 & 99.7 & 99.0 & 99.5 \\ 
 & FPR & 4.2 & 0.4 & 7.2 & 7.7 & 0.8 & 0.3 & 14.0 & 0.5 & 1.0 \\

    \end{tabular}}

    \label{tab:gender-shades}
     \vspace{-3mm}
\end{table}

\section{Learning inclusion probabilities}
\label{sec:debiasing}

In social sensing studies, the results of measurements performed on a given platform, e.g., a social media site, are studied to understand the behavior of a population. Often, such measurements are biased~\cite{ruths2014social}, as individuals with certain demographics are more likely to join these platforms, e.g., young people may be more likely to join Twitter.
Obtaining representative estimates in these scenarios is challenging, as the probability of an individual with given demographics to be on a given platform, also referred to as \textit{inclusion probability} in sampling methodology, is typically unknown. 

Here, we estimate these inclusion probabilities as a function of demographics. To this end, we learn the debiasing coefficients on the grounds of statistical survey analysis with missing data~\cite{bethlehem1987linear,Little2002,Sarndal2005}. Specifically, we derive estimators for the numbers of individuals by making an assumption that per-strata inclusion probabilities are equal for individuals within a group (e.g., a country) and different between groups (e.g., between countries). 
This assumption allows us to get both across-groups global debiasing coefficients (i.e., the biases of Twitter users) and the group-specific debiasing coefficients (i.e., per-country demographics on Twitter, e.g., developed countries may have more and diverse users on Twitter). 
Our approach learns the inclusion probabilities, while typical post-stratification methods either assume these probabilities are known or focus on obtaining a post-stratified estimator of a response variable.

\subsection{Formulation of Debiasing Models}

Consider a population $U$ of $N=|U|$ individuals with certain demographics.
For simplicity, we focus on the case of only two discrete demographic variables, say age $a$ and gender $g$, which are distributed following $P_\text{N}(a,g)$, but the following reasoning applies to other demographics as well. 
Out of these $N$ individuals, say, $M \leq N$ joined a certain online platform with a probability depending on their demographics. For instance, individuals joined Twitter and younger people were more likely to join than older ones. 

In survey analysis, the probability that an individual with certain demographics joined a certain platform corresponds to the inclusion probability of the stratum representing these demographics \cite{bethlehem1987linear, Sarndal2005}. In the simplest scenario, when the probability of joining the platform is homogeneous in time and across individuals, this probability can be expressed as the ratio of the number of Twitter users with the demographics $a$ and $g$ to all individuals, i.e., $\pi (a,g) = \frac{M(a,g)}{N(a,g)} = \frac{M P_\text{M}(a,g)}{N P_\text{N}(a,g)}$, where $P_\text{M}(a,g)$ and $P_\text{N}(a,g)$ are the distributions of demographics of Twitter users and the overall population, respectively. 
However, the inclusion probability may vary between individuals. 
To account for this, we discuss the homogeneity of inclusion probability for a given partition of the population. A partition of the population $U$ is defined through non-overlapping and non-empty subsets $U_i$ of the population that together constitute $U$, i.e., $\bigcup_i U_i = U$.  
Typically such subsets will have a certain meaning, e.g., a natural partition of a population is a split by countries, regions, or cities.
The total number of individuals in the subset $i$ is $N^i$ and the number of individuals having particular demographics is $N^i(a,g)$.

\subsubsection{Homogeneous bias}

The inclusion probability, which governs the bias, is homogeneous with respect to a given set of demographics and a given partition of the population, if and only if $\pi^i (a,g) = \pi(a,g)$ for each subset $i$ of the partition, i.e., the inclusion probability does not depend on the elements of the partition. 

If the inclusion probability is homogeneous, then we can write
\begin{align}
N^i &= \sum_a \sum_g N^i(a,g) 
= \sum_a \sum_g \frac { M^i(a,g)} {\pi(a,g)}.
\label{eq:model1}
\end{align}
Thus, to obtain the inclusion probabilities, we regress $N^i$ against $M^i(a,g)$. If the condition of homogeneity holds, then the regression coefficients are equal to $1/\pi(a,g)$.

\subsubsection{Inhomogeneous bias}

If the inclusion probability is inhomogenous for given demographics and partition, then we shall model each subset of the partition separately or use a different model that relaxes the homogenity assumption by specifying the functional form of inhomogeneity. Here, we consider the inhomogeneity of the form 
\begin{equation}
\pi^i (a,g) = M^i(a,g)^\nu f_1(a) f_2(g),
\label{eq:inhomogen}
\end{equation}
where $\nu$ is an unknown exponent and $f_1$ and $f_2$ are unknown functions.\footnote{Note that the aforementioned homogeneity assumption is a special case of this assumption, i.e., this assumption relaxes the homogeneity assumption.} 
Under this assumption,
$\log N^i(a,g) = \log \frac{M^i(a,g)}{\pi^i(a,g)} = (1-\nu) \log M^i(a,g) - \log f_1(a) - \log f_2(g).$
This time, to obtain the debiasing coefficients and corresponding inclusion probabilities, we regress the ground-truth population, $\log N^i(a,g)$, against the biased measurements, $\log M^i(a,g)$, and the demographic indicator variables, i.e.,
\begin{align}
\log N^i(a,g) &= \beta_1 \log M^i(a,g) + 
\sum_{\tilde{a}} \beta_{\tilde{a}}\delta_{\tilde{a}a} +
\sum_{\tilde{g}} \beta_{\tilde{g}}\delta_{\tilde{g}g}
\label{eq:model2}
\end{align}
where $\delta_{ij}$ is Kronecker delta, i.e., $\delta _{ij}=0$ if $i\neq j$ and $\delta _{ij}=1$ if $i=j$, $\beta_1=(1-v)$, $\beta_{a}=-\log \ f_1(a)$, and $\beta_{g}=-\log \ f_2(g)$. From these regression coefficients we can obtain the inclusion probability $\pi^i (a,g)$ of a set of samples $i$ with the demographics $a$ and $g$ via Equation~\ref{eq:inhomogen}. 
This debiasing model has been proposed recently by Zagheni et al. to predict the number of migrants based on Facebook data~\cite{Zagheni2017Leveraging}, but its derivation and explicit formal interpretation have not been provided until now, to the best of our knowledge.
Note that this estimation method of inclusion probabilities requires the ground-truth joint counts $N^i(a,g)$ at the time of training, whereas the model based on the homogeneity assumption requires only the total counts $N^i$. The availability of the joint counts is often limited in practice because of insufficient number of samples per stratum.

\subsection{Discussion}

We derived debiasing models by making homogeneity or inhomogeneity assumptions. These models predict population size, so they can be evaluated by measuring the error of predictions in cross-validation settings or via model selection methods. In this way, we test which assumption is closer to reality and find the most accurate method for the given dataset. We evaluate these models in the next section using the population  of  regions in EU countries. %

\section{European Population Inference from Twitter data}
\label{sec:twitter}

Here, we use the debiasing models formulated in the previous section to obtain the debiasing coefficients and the corresponding inclusion probabilities for Twitter users of different countries. To this end, we regress the country-level ground-truth number of people living in a certain location against the number of Twitter users of different demographics. 

\subsection{Debiasing Models for Population Inference}

More specifically, we evaluate five models requiring different amounts of data. The first model is a baseline, the next three models are based on the assumption of homogeneous inclusion probabilities (Equation~\ref{eq:model1}), whereas the last one is based on the inhomogeneity assumption (Equation~\ref{eq:model2}):

\def\bestmodel{M(a,g) + a+g}
\begin{description}
\item[$\mathbf{N \sim M}$] is our base model that uses only the total population count from the census ($N$) and Twitter ($M$).
\item[$\mathbf{N \sim \sum_g M(g)}$] uses gender marginal counts only (i.e., the total counts of males and females not broken down by ages).
\item[$\mathbf{N \sim \sum_a M(a)}$] uses age marginal counts only.
\item[$\mathbf{N \sim \sum_{a,g} M(a,g)}$] uses the joint histograms inferred from Twitter but only the total population values from the census.
\item[$\mathbf{\text{\textbf{log}}~N(a,g)  \sim \text{\textbf{log}}~\bestmodel}$] uses the joint histograms inferred from Twitter and the joint histograms from the census.
\end{description}

Note that Twitter users are biased in various ways: the platform is more accessible to tech-savvy individuals and citizens of various countries use it to a different extent \cite{hale2014global}.
To distinguish the global effect of Twitter on the overall bias from the local effect of country, we use multilevel models. Namely, all slopes and intercepts in the introduced models have random effects specific to a country.
Thus, from fixed effects of the model, we obtain global debiasing coefficients (i.e., the biases of Twitter users) and, from random effects, the country-specific debiasing coefficients (i.e., a given country has its own bias towards Twitter). %

Note that the homogeneity and inhomogeneity assumptions apply within each country separately. Namely, we group samples by regions of a country, described next. %
Then, the homogeneity assumption translates into the same inclusion probability for a given demographic across all regions of a country, whereas the inhomogeneity assumption translates into a dependence of the inclusion probability on regions of a country that follows Equation~\ref{eq:inhomogen}.

\subsection{Data}

We retrieved joint population distributions for age and gender on a regional level (NUTS3) for 26 countries of the European Union plus four EFTA members as made available through the CensusHub system of Eurostat, the European Statistical Office.\footnote{\url{https://ec.europa.eu/CensusHub2/}: Luxembourg and Belgium were not available. Additional EFTA countries: Switzerland, Iceland, Liechtenstein and Norway.} All census data is from 2011, the most recent year for which comprehensive census data was collectively reported to Eurostat. NUTS3 regions are the finest-grained level of the ``Nomenclature of Territorial Units for Statistics'' used as a standard for statistical reporting in EU member states~\cite{eunuts3}. They are based on existing local administrative boundaries and are usually at the level of local districts.
Although sizes differ to a certain extent per country, NUTS3 provide the most standardized cross-country geographical units to subdivide populations. We use the 2010 iteration of the NUTS3 regions as these correspond with the 2011 census subdivisons and is the most recent census available.

Our sample of social media users is derived from the random 10\% stream of tweets from Twitter. We recorded all users observed from September 2015 to January 2016 and downloaded profile photos for all users in spring 2018.  The location of each user is inferred using the method of Compton et al.~\cite{compton2014geotagging}.  This model predicts a user's latitude and longitude and was shown to be least biased with respect to urban and rural areas \cite{johnson2017effect}, which is important given the diversity in population centers in our study.  
The model was trained on a social network of 781M edges constructed from Twitter data spanning 2012 to 2017.  Five-fold cross validation reports a median inference error of 7.9km, which is sufficiently accurate for the geographic granularity we use here.  %
Each user is assigned to a NUTS3 region in Europe if the inferred latitude-longitude pair of that user lies within the boundaries of the NUTS3 regions; users not in these regions are discarded.
Ultimately, we obtain a dataset of 3,202,964 users within the NUTS3 regions for our study. In the remainder, we will refer to NUTS3 regions simply as ``regions.''

The age, gender, and is-organization variables for each user in our dataset are inferred using M3.  As an additional experiment to quantify the impact of including non-human accounts, we ignore any organization classification and treat all accounts as humans, grouping them according to their inferred demographics.

\subsection{Results}

We evaluate the debiasing models in the following cross-validation settings: leave one region out, leave one country out (i.e., leave out all regions from a given country), and leave one stratum out (e.g., leave out only males aged 30-39).
Per case, we measure the mean absolute percentage error of the population estimates for the left-out samples, i.e., $ \text{MAPE}(N) = \frac{100\%}{n} \sum_i \frac{ |\hat{N}_i-N_i|}{|N_i|} $,
where $\hat{N}_i$ and $N_i$ are the predicted and actual population sizes, respectively, and the sum is over all regions. Note that the model \mbox{$\log N(a,g) \sim \log \bestmodel$} operates naturally in the log space; hence, before calculating the error of this model we first exponentiate the predicted population sizes to make the results more comparable across all models. Its results should further be compared to the remaining models with care, as the effects of moving to log space cannot be clearly untangled from gains due to learning on joint age and gender attributes.
\def\mZeroMAPE{88\%}
\def\mMGenderMAPE{59\%}
\def\mMAgeMAPE{61\%}
\def\mOneMAPE{54\%}
\def\mBTTMAPE{33\%}

\begin{figure}
    \centering
    \includegraphics[width=0.45\textwidth]{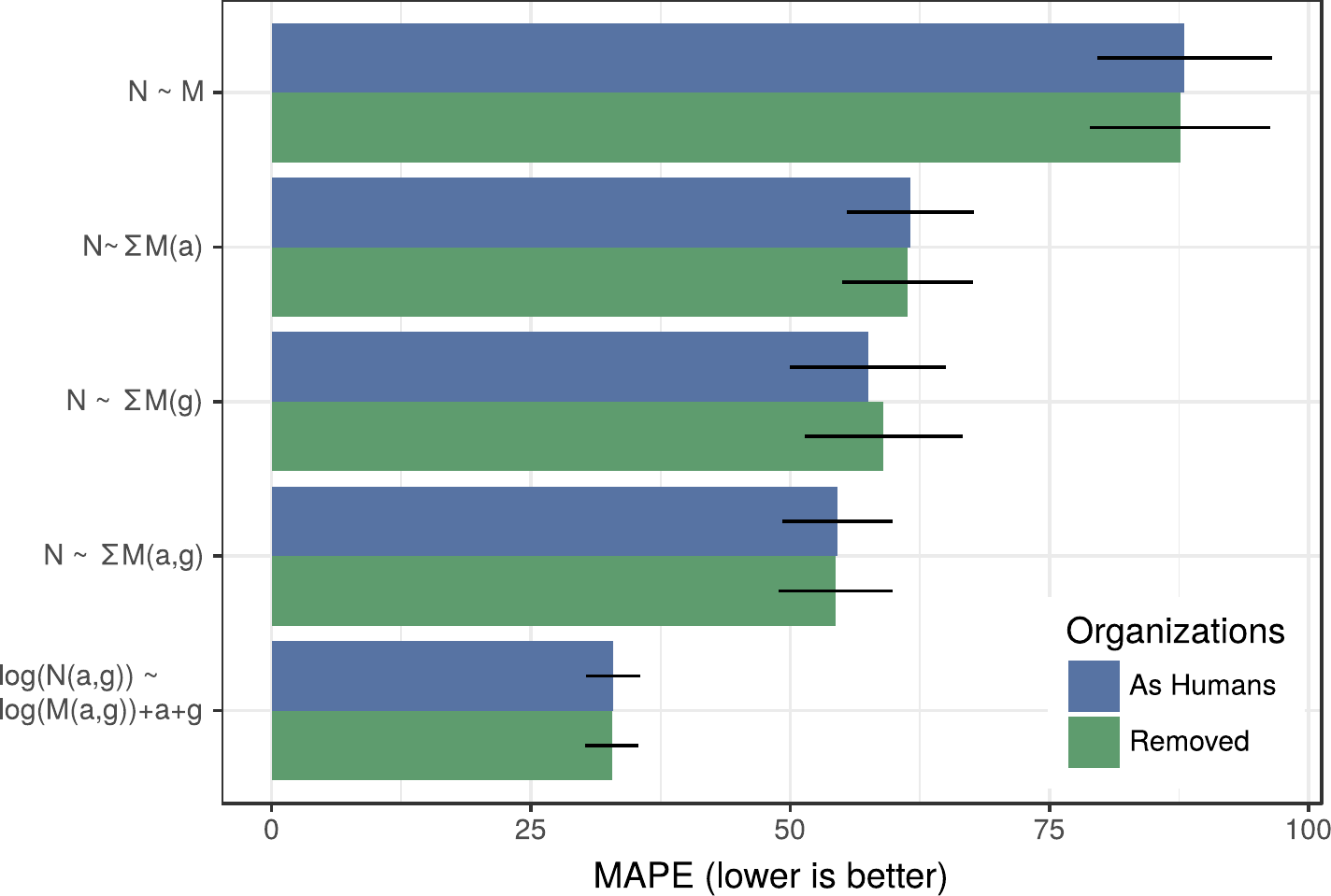}
    \vspace{-2mm}
    \caption{Performance on leave-one-region-out population inference across different debiasing models. Bars show mean MAPE($N$) with 95\% confidence intervals. The results of the last model are exponentiated to compute MAPE($N$).}
    \label{fig:debiasing-performance}
    \vspace{-3mm}
\end{figure}

The results of the leave-one-region-out evaluation show a clear benefit to debiasing based on the inferred demographics (Figure~\ref{fig:debiasing-performance}). For the $N \sim M$ model without demographics, MAPE is \mZeroMAPE{}. The inclusion of the inferred gender or age demographics in the debiasing models, via $N \sim \sum M(a)$ and $N \sim \sum M(g)$, decreases MAPE to \mMGenderMAPE{} and \mMAgeMAPE{}, respectively. %
Including inferred joint distributions, e.g., the number of males of age 30-39, in the model $N \sim \sum M(a,g)$, decreases MAPE further to \mOneMAPE{}, even though we do not use joint-distribution data from the census for training. In fact, the use of census joint distributions at the stage of model training along with a move to log space, via $\log N(a,g) \sim \log \bestmodel$, further improves the accuracy, bringing the MAPE down to \mBTTMAPE{}. 
To understand the effect of accounts belonging to organizations on this population prediction task, we compare the results of the debiasing models with and without removing the organizations (Figure~\ref{fig:debiasing-performance}). Removing organizations results in a small reduction in error for all but the model trained with only marginal gender counts, though these differences are not statistically significant. A further analysis shows that the presence of organizations is heavily skewed towards populous metropolitan cities and most regions have very few organizations (i.e., organizations are not equally geographically distributed). However, for populous regions, organizations can create significant error.

\begin{figure}
    \centering
    \includegraphics[width=0.5\textwidth]{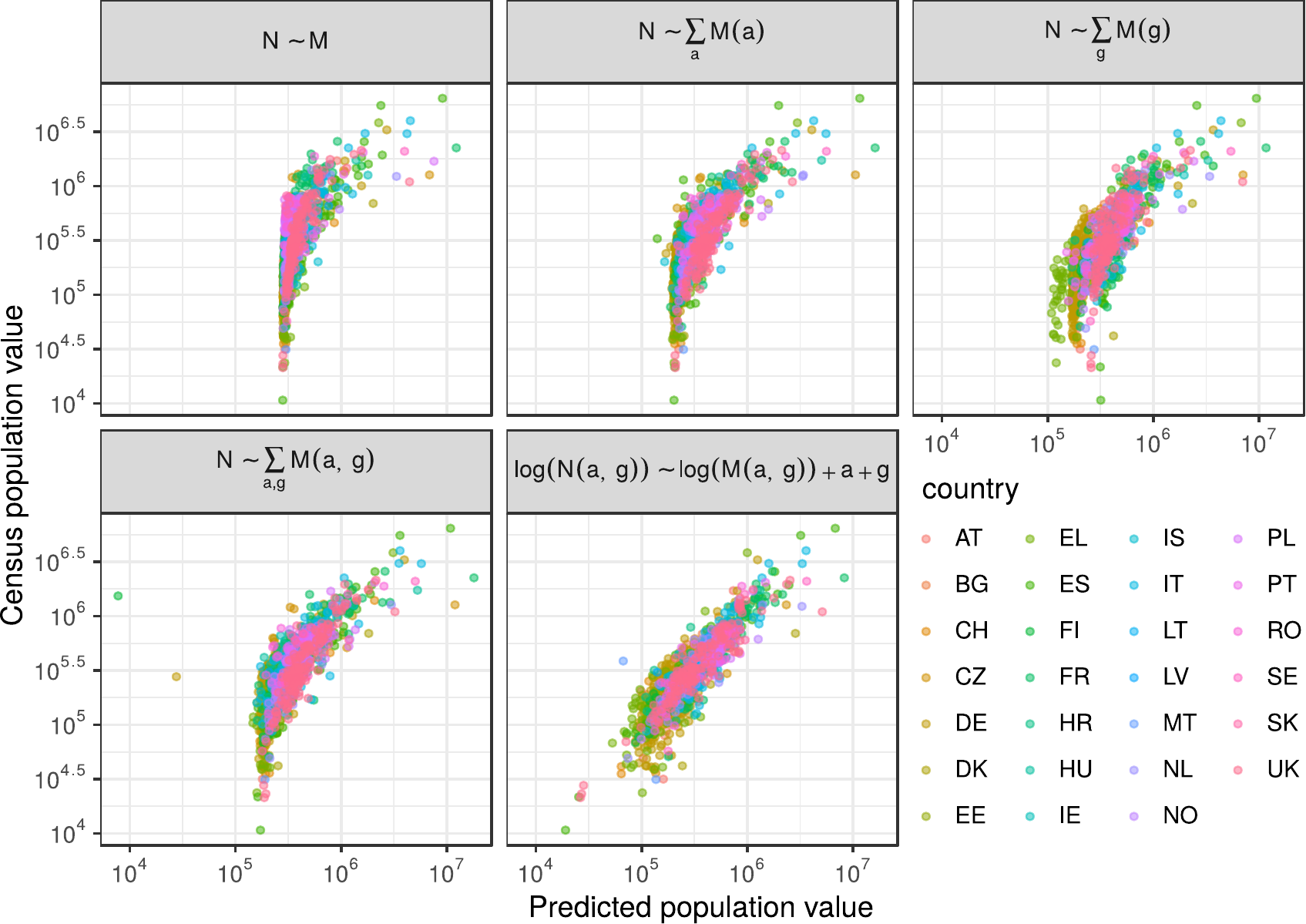}
    \caption{Comparing the census population and debiased estimates for leave-one-region-out evaluations shows that the model with joint Twitter and census distributions correlates best with the real values for large and small regions.}
    \label{fig:scatter}
    \vspace{-3mm}
\end{figure}

\begin{figure}
    \centering
    \includegraphics[width=0.5\textwidth]{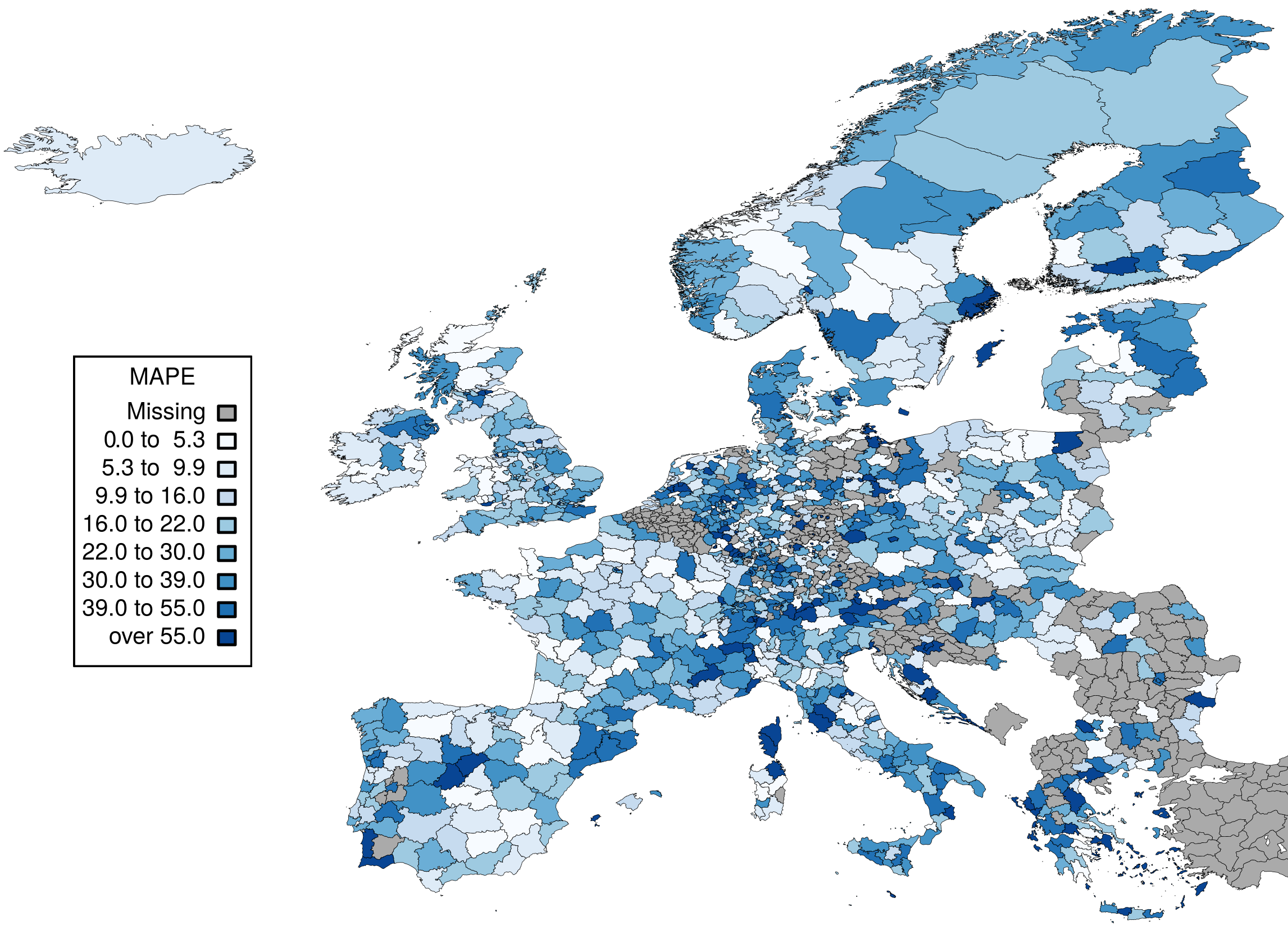}
    \caption{MAPE  for $\log N(a,g) \sim \log \bestmodel$ model across regions shows that many regions have population estimate errors below 10\% of the true population, with high-error regions found infrequently in all countries. For greyed regions we either miss NUT3 data or Twitter counts are zero.}
    \label{fig:mape_map}
\end{figure}

To gain further insights into the introduced models, we show scatter plots of true and predicted population sizes for each model (Figure~\ref{fig:scatter}). The model with joint Twitter and census counts is noticeably closer to the $y=x$ line, likely because this models is trained in log space. For this model, we also depict the geographical variation of MAPE across the EU regions (Figure \ref{fig:mape_map}).

\begin{figure}
    \centering
    \includegraphics[width=0.33\textwidth]{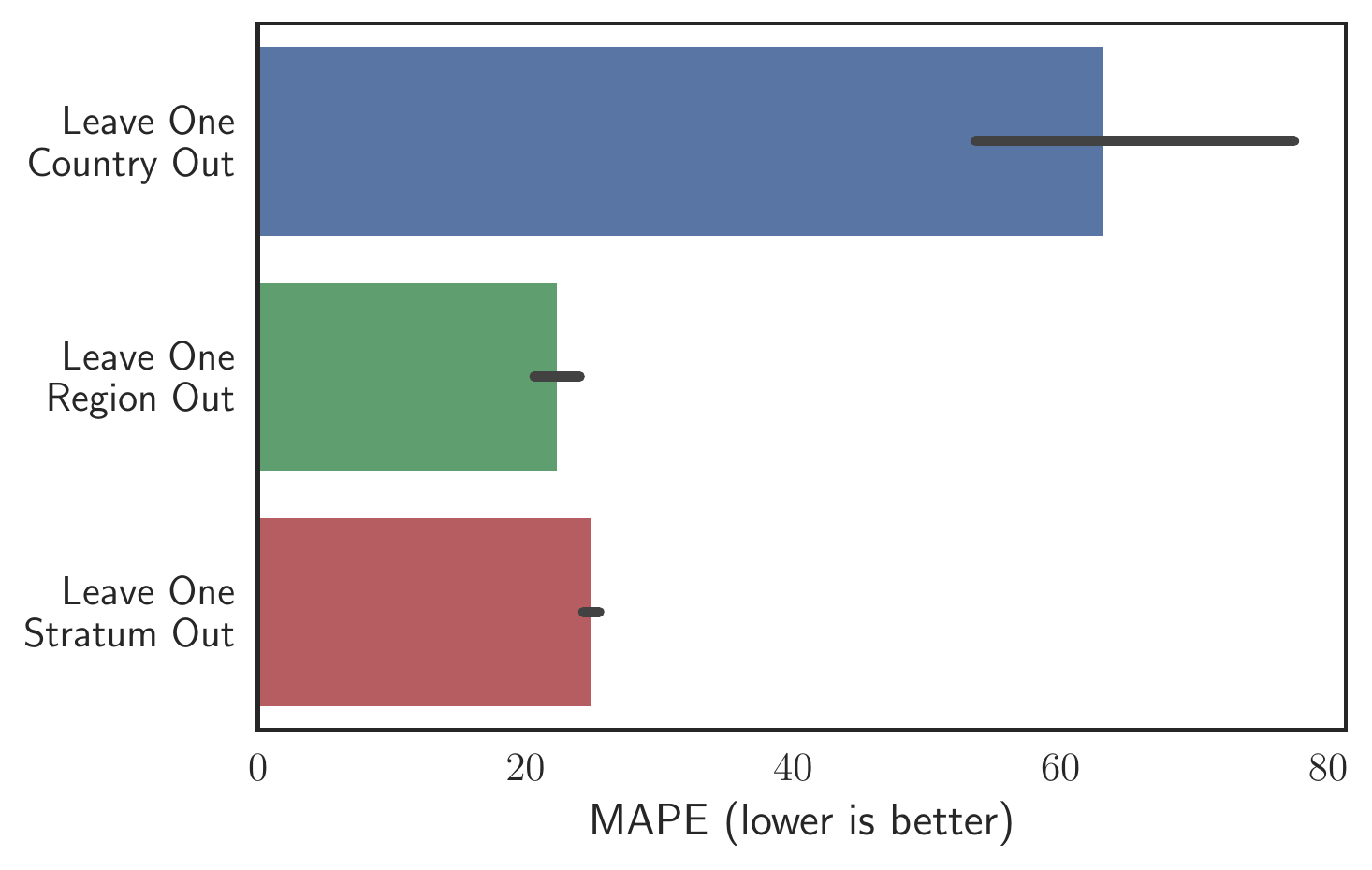}
    \caption{Comparisons of the model with joint Twitter and census distributions as increasing amounts of information are held out from training: (i) leaving out one stratum, (ii) leaving out all the strata from one region, and (iii) leaving out all regions in one country.}
    \label{fig:different-eval-plot}
    \vspace{-4mm}
\end{figure}

The above results test predictions when one region is left out and its population is predicted. However, in different circumstances, more or less census information could be available. We test two other cross-validation settings: leave one stratum out (e.g., leave out only females aged 30--39) and leave one country out (i.e., leave out all regions from a given country). The latter reflects the generalizability of the model to completely unseen countries where platform adoption probabilities and country-specific biases are not known. The evaluation results for all three cross-validation settings are plotted in Figure \ref{fig:different-eval-plot} for the most accurate model, i.e., $\log N(a,g) \sim \log \bestmodel$. From the first additional cross-validation setting, we learn that hiding the population sizes  for a specific stratum in all regions   results in  a minimal penalty to the prediction accuracy. For the second additional cross-validation setting, we see that the error rate nearly 
doubles with an average MAPE of 81\%.
This result suggests that knowledge of country-specific platform biases is important for accurate estimates and that at least some regions within the country should be seen during a model's training time to reach high performance.

\subsection{Discussion of Debiasing Results and Potential Sources of Error}
Debiasing social media data samples is a difficult task, but our results show that automatic post-stratification with respect to inferred age and gender notably improves population estimation results. Furthermore, our predictive models are fully interpretable, which allows us to estimate the inclusion probabilities and share them with the research community for future reuse. 

Even when joint distributions are not given by the census, inferring joint distributions from social media data with the model $N \sim M(a,g)$ provides a significant increase in accuracy in population prediction tasks compared to the baseline without debiasing. However, the model $\log N(a,g) \sim \log \bestmodel$ is notably more accurate, suggesting that the assumption of homogeneity does not hold for the partition of citizens of a particular country, gender, and age into regions. Note, however, that the predictions of the most accurate model are not perfect. This may be caused by the time mismatch of about 5 years between the Twitter and NUTS3 datasets. On the other hand, this point opens the door for developing debiasing models based on other inhomogeneity assumptions and searching better partitions, e.g., our multilevel model could have more levels to capture biases shared by smaller regions within a country.

\begin{figure}
    \centering
    \includegraphics[width=0.5\textwidth]{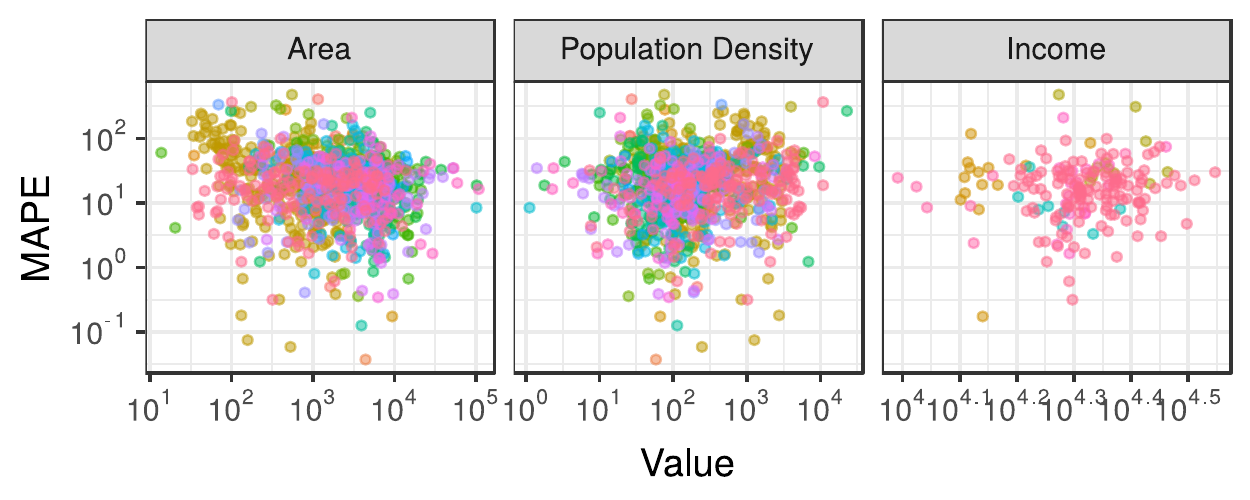}
    \vspace{-6mm}
    \caption{Comparing the model with joint Twitter and census distributions MAPE to region area (km$^2$), population density (people/km$^2$), and income (USD per capita) shows that the model performance is not biased towards particular types of regions. Colors represent countries as in Figure~\ref{fig:scatter}.}
    \label{fig:area_density_income}
    \vspace{-4mm}
\end{figure}

To learn more about potential confounders, we compare the MAPE of the model $\log N(a,g)  \sim \log \bestmodel$ to three variables: the area of the region, its population density, and average income (Figure \ref{fig:area_density_income}).
NUTS3 regions are defined by each country and vary considerably in land area from city-states in Germany to large regions in Sweden (up to 98,000 km$^2$). We do not find correlations of land area  with the MAPE error.
Hecht and Stephens \cite{hecht2014urban} found a clear urban--rural bias in social media data: there tend to be more users, more information, and higher quality information per capita within metropolitan areas. Despite this bias, we find our estimates not heavily correlated with population density,  suggesting our models do well at debiasing these differences in inclusion probabilities. Only two countries (Czech Republic and Norway) show a correlation that is statically different from zero across our models. %
Finally, income has been suggested as an important explanatory variable for who uses Internet-based platforms and hence could be a potential confound to our estimates \cite{chinn2007determinants,jansen2010use}. However, we similarly find that income and MAPE are not correlated significantly in our models.

\section{Conclusion}

The everyday opinions expressed in social media provide promising opportunities for measuring population-level statistics for health metrics, political outcomes, or general attitudes.  However, social media is a non-representative sample of the population due to demographic skew in usage frequencies and access rates.  As such, any direct estimate from a platform is likely biased towards certain demographics.  This work provides a holistic solution to this problem by developing a novel method for assigning users to demographic strata and exploiting it in a regression framework for debiasing that allows direct estimation of the probability of an individual with given demographics to be on the given social media platform.  Our work provides three main contributions. First, we introduce a state-of-the-art  neural system for multi-attribute classification in 32 languages.  This contribution also includes the system release and the creation of a new dataset of gender, age, and is-organization annotations in 32 languages. Second, we derive a series of models to debias social media measurements and provide their explicit formal interpretations. Third, in a massive study of all of Europe, we show that our two methods are able to infer regional population counts accurately and provide demographic corrections for all downstream measurements on the grounds of the estimated inclusion probabilities. These results pave the way for more accurate social sensing by laying a foundation of representative population sampling in social media.  Code, software, and debiasing coefficients  pertaining to this work are released for public use at {\small \url{https://github.com/euagendas/}}. 

\section*{Acknowledgements}
This research has received funding through the Volkswagen Foundation and was supported by The Alan Turing Institute under EPSRC grant EP/N510129/1. %

\bibliographystyle{ACM-Reference-Format}
\balance
\bibliography{small-references}


\begin{thebibliography}{82}


\ifx \showCODEN    \undefined \def \showCODEN     #1{\unskip}     \fi
\ifx \showDOI      \undefined \def \showDOI       #1{#1}\fi
\ifx \showISBNx    \undefined \def \showISBNx     #1{\unskip}     \fi
\ifx \showISBNxiii \undefined \def \showISBNxiii  #1{\unskip}     \fi
\ifx \showISSN     \undefined \def \showISSN      #1{\unskip}     \fi
\ifx \showLCCN     \undefined \def \showLCCN      #1{\unskip}     \fi
\ifx \shownote     \undefined \def \shownote      #1{#1}          \fi
\ifx \showarticletitle \undefined \def \showarticletitle #1{#1}   \fi
\ifx \showURL      \undefined \def \showURL       {\relax}        \fi
\providecommand\bibfield[2]{#2}
\providecommand\bibinfo[2]{#2}
\providecommand\natexlab[1]{#1}
\providecommand\showeprint[2][]{arXiv:#2}

\bibitem[\protect\citeauthoryear{Al~Zamal, Liu, and Ruths}{Al~Zamal
  et~al\mbox{.}}{2012}]%
        {al2012homophily}
\bibfield{author}{\bibinfo{person}{Faiyaz Al~Zamal}, \bibinfo{person}{Wendy
  Liu}, {and} \bibinfo{person}{Derek Ruths}.} \bibinfo{year}{2012}\natexlab{}.
\newblock \showarticletitle{Homophily and Latent Attribute Inference: Inferring
  Latent Attributes of Twitter Users from Neighbors}. In
  \bibinfo{booktitle}{\emph{Proceedings of ICWSM}}.
\newblock


\bibitem[\protect\citeauthoryear{Alzahrani, Gore, Salehi, and
  Davulcu}{Alzahrani et~al\mbox{.}}{2018}]%
        {alzahrani2018finding}
\bibfield{author}{\bibinfo{person}{Sultan Alzahrani}, \bibinfo{person}{Chinmay
  Gore}, \bibinfo{person}{Amin Salehi}, {and} \bibinfo{person}{Hasan Davulcu}.}
  \bibinfo{year}{2018}\natexlab{}.
\newblock \showarticletitle{Finding Organizational Accounts Based on Structural
  and Behavioral Factors on Twitter}. In
  \bibinfo{booktitle}{\emph{International Conference on Social Computing,
  Behavioral-Cultural Modeling and Prediction and Behavior Representation in
  Modeling and Simulation}}. Springer, \bibinfo{pages}{164--175}.
\newblock


\bibitem[\protect\citeauthoryear{An and Weber}{An and Weber}{2015}]%
        {an2015whom}
\bibfield{author}{\bibinfo{person}{Jisun An} {and} \bibinfo{person}{Ingmar
  Weber}.} \bibinfo{year}{2015}\natexlab{}.
\newblock \showarticletitle{Whom should we sense in ``social
  sensing''---Analyzing which users work best for social media now-casting}.
\newblock \bibinfo{journal}{\emph{EPJ Data Science}} \bibinfo{volume}{4},
  \bibinfo{number}{1} (\bibinfo{date}{30 Nov} \bibinfo{year}{2015}),
  \bibinfo{pages}{22}.
\newblock
\showISSN{2193-1127}
\urldef\tempurl%
\url{https://doi.org/10.1140/epjds/s13688-015-0058-9}
\showDOI{\tempurl}


\bibitem[\protect\citeauthoryear{Ardehaly and Culotta}{Ardehaly and
  Culotta}{2017}]%
        {ardehaly2017co}
\bibfield{author}{\bibinfo{person}{Ehsan~Mohammady Ardehaly} {and}
  \bibinfo{person}{Aron Culotta}.} \bibinfo{year}{2017}\natexlab{}.
\newblock \showarticletitle{Co-training for Demographic Classification Using
  Deep Learning from Label Proportions}. In \bibinfo{booktitle}{\emph{Data
  Mining Workshops (ICDMW), 2017 IEEE International Conference on}}. IEEE,
  \bibinfo{pages}{1017--1024}.
\newblock


\bibitem[\protect\citeauthoryear{Azure}{Azure}{2018}]%
        {microsoft18face}
\bibfield{author}{\bibinfo{person}{Microsoft Azure}.}
  \bibinfo{year}{2018}\natexlab{}.
\newblock \bibinfo{title}{Cognitive Services}.
\newblock
\newblock
\urldef\tempurl%
\url{https://azure.microsoft.com/en-us/services/cognitive-services/}
\showURL{%
\tempurl}


\bibitem[\protect\citeauthoryear{Bamman, Eisenstein, and Schnoebelen}{Bamman
  et~al\mbox{.}}{2014}]%
        {bamman2014gender}
\bibfield{author}{\bibinfo{person}{David Bamman}, \bibinfo{person}{Jacob
  Eisenstein}, {and} \bibinfo{person}{Tyler Schnoebelen}.}
  \bibinfo{year}{2014}\natexlab{}.
\newblock \showarticletitle{Gender identity and lexical variation in social
  media}.
\newblock \bibinfo{journal}{\emph{Journal of Sociolinguistics}}
  \bibinfo{volume}{18}, \bibinfo{number}{2} (\bibinfo{year}{2014}),
  \bibinfo{pages}{135--160}.
\newblock


\bibitem[\protect\citeauthoryear{Barchiesi, Moat, Alis, Bishop, and
  Preis}{Barchiesi et~al\mbox{.}}{2015}]%
        {Barchiesi2015Quantifying}
\bibfield{author}{\bibinfo{person}{Daniele Barchiesi},
  \bibinfo{person}{Helen~Susannah Moat}, \bibinfo{person}{Christian Alis},
  \bibinfo{person}{Steven Bishop}, {and} \bibinfo{person}{Tobias Preis}.}
  \bibinfo{year}{2015}\natexlab{}.
\newblock \showarticletitle{Quantifying International Travel Flows Using
  Flickr}.
\newblock \bibinfo{journal}{\emph{PLOS ONE}} \bibinfo{volume}{10},
  \bibinfo{number}{7} (\bibinfo{date}{07} \bibinfo{year}{2015}),
  \bibinfo{pages}{1--8}.
\newblock
\urldef\tempurl%
\url{https://doi.org/10.1371/journal.pone.0128470}
\showDOI{\tempurl}


\bibitem[\protect\citeauthoryear{Beller, Knowles, Harman, Bergsma, Mitchell,
  and Van~Durme}{Beller et~al\mbox{.}}{2014}]%
        {beller2014ma}
\bibfield{author}{\bibinfo{person}{Charley Beller}, \bibinfo{person}{Rebecca
  Knowles}, \bibinfo{person}{Craig Harman}, \bibinfo{person}{Shane Bergsma},
  \bibinfo{person}{Margaret Mitchell}, {and} \bibinfo{person}{Benjamin
  Van~Durme}.} \bibinfo{year}{2014}\natexlab{}.
\newblock \showarticletitle{I'm a belieber: Social roles via
  self-identification and conceptual attributes}. In
  \bibinfo{booktitle}{\emph{Proceedings of the 52nd Annual Meeting of the
  Association for Computational Linguistics (Volume 2: Short Papers)}},
  Vol.~\bibinfo{volume}{2}. \bibinfo{pages}{181--186}.
\newblock


\bibitem[\protect\citeauthoryear{Bergsma and Van~Durme}{Bergsma and
  Van~Durme}{2013}]%
        {bergsma2013using}
\bibfield{author}{\bibinfo{person}{Shane Bergsma} {and}
  \bibinfo{person}{Benjamin Van~Durme}.} \bibinfo{year}{2013}\natexlab{}.
\newblock \showarticletitle{Using Conceptual Class Attributes to Characterize
  Social Media Users}. In \bibinfo{booktitle}{\emph{Proc. ACL}}.
\newblock


\bibitem[\protect\citeauthoryear{Bethlehem and Keller}{Bethlehem and
  Keller}{1987}]%
        {bethlehem1987linear}
\bibfield{author}{\bibinfo{person}{Jelke~G Bethlehem} {and}
  \bibinfo{person}{Wouter~J Keller}.} \bibinfo{year}{1987}\natexlab{}.
\newblock \showarticletitle{Linear weighting of sample survey data}.
\newblock \bibinfo{journal}{\emph{Journal of Official Statistics}}
  \bibinfo{volume}{3}, \bibinfo{number}{2} (\bibinfo{year}{1987}),
  \bibinfo{pages}{141--153}.
\newblock


\bibitem[\protect\citeauthoryear{Blum and Mitchell}{Blum and Mitchell}{1998}]%
        {blum1998combining}
\bibfield{author}{\bibinfo{person}{Avrim Blum} {and} \bibinfo{person}{Tom
  Mitchell}.} \bibinfo{year}{1998}\natexlab{}.
\newblock \showarticletitle{Combining labeled and unlabeled data with
  co-training}. In \bibinfo{booktitle}{\emph{Proceedings of the eleventh annual
  conference on Computational learning theory}}. ACM, \bibinfo{pages}{92--100}.
\newblock


\bibitem[\protect\citeauthoryear{Buolamwini and Gebru}{Buolamwini and
  Gebru}{2018}]%
        {buolamwini2018gender}
\bibfield{author}{\bibinfo{person}{Joy Buolamwini} {and}
  \bibinfo{person}{Timnit Gebru}.} \bibinfo{year}{2018}\natexlab{}.
\newblock \showarticletitle{Gender shades: Intersectional accuracy disparities
  in commercial gender classification}. In \bibinfo{booktitle}{\emph{Conference
  on Fairness, Accountability and Transparency}}. \bibinfo{pages}{77--91}.
\newblock


\bibitem[\protect\citeauthoryear{Chen, Wang, Agichtein, and Wang}{Chen
  et~al\mbox{.}}{2015}]%
        {chen2015comparative}
\bibfield{author}{\bibinfo{person}{Xin Chen}, \bibinfo{person}{Yu Wang},
  \bibinfo{person}{Eugene Agichtein}, {and} \bibinfo{person}{Fusheng Wang}.}
  \bibinfo{year}{2015}\natexlab{}.
\newblock \showarticletitle{A Comparative Study of Demographic Attribute
  Inference in Twitter.}. In \bibinfo{booktitle}{\emph{Proceedings of ICWSM}},
  Vol.~\bibinfo{volume}{15}. \bibinfo{pages}{590--593}.
\newblock


\bibitem[\protect\citeauthoryear{Chinn and Fairlie}{Chinn and Fairlie}{2007}]%
        {chinn2007determinants}
\bibfield{author}{\bibinfo{person}{Menzie~D Chinn} {and}
  \bibinfo{person}{Robert~W Fairlie}.} \bibinfo{year}{2007}\natexlab{}.
\newblock \showarticletitle{The determinants of the global digital divide: a
  cross-country analysis of computer and internet penetration}.
\newblock \bibinfo{journal}{\emph{Oxford Economic Papers}}
  \bibinfo{volume}{59}, \bibinfo{number}{1} (\bibinfo{year}{2007}),
  \bibinfo{pages}{16--44}.
\newblock


\bibitem[\protect\citeauthoryear{Chung, Cho, and Bengio}{Chung
  et~al\mbox{.}}{2016}]%
        {chung2016character}
\bibfield{author}{\bibinfo{person}{Junyoung Chung}, \bibinfo{person}{Kyunghyun
  Cho}, {and} \bibinfo{person}{Yoshua Bengio}.}
  \bibinfo{year}{2016}\natexlab{}.
\newblock \showarticletitle{A character-level decoder without explicit
  segmentation for neural machine translation}.
\newblock In \bibinfo{booktitle}{\emph{Proceedings of the 54th Annual Meeting
  of the Association for Computational Linguistics}}.
  \bibinfo{publisher}{Association for Computational Linguistics},
  \bibinfo{pages}{1693--1703}.
\newblock


\bibitem[\protect\citeauthoryear{Ciot, Sonderegger, and Ruths}{Ciot
  et~al\mbox{.}}{2013}]%
        {ciot2013gender}
\bibfield{author}{\bibinfo{person}{Morgane Ciot}, \bibinfo{person}{Morgan
  Sonderegger}, {and} \bibinfo{person}{Derek Ruths}.}
  \bibinfo{year}{2013}\natexlab{}.
\newblock \showarticletitle{Gender inference of Twitter users in non-English
  contexts}. In \bibinfo{booktitle}{\emph{Proceedings of the 2013 Conference on
  Empirical Methods in Natural Language Processing}}.
  \bibinfo{pages}{1136--1145}.
\newblock


\bibitem[\protect\citeauthoryear{Coates}{Coates}{1998}]%
        {coates1998language}
\bibfield{author}{\bibinfo{person}{Jennifer Coates}.}
  \bibinfo{year}{1998}\natexlab{}.
\newblock \bibinfo{booktitle}{\emph{Language and gender: A reader}}.
\newblock \bibinfo{publisher}{Wiley-Blackwell}.
\newblock


\bibitem[\protect\citeauthoryear{Coates}{Coates}{2015}]%
        {coates2015women}
\bibfield{author}{\bibinfo{person}{Jennifer Coates}.}
  \bibinfo{year}{2015}\natexlab{}.
\newblock \bibinfo{booktitle}{\emph{Women, men and language: A sociolinguistic
  account of gender differences in language}}.
\newblock \bibinfo{publisher}{Routledge}.
\newblock


\bibitem[\protect\citeauthoryear{Compton, Jurgens, and Allen}{Compton
  et~al\mbox{.}}{2014}]%
        {compton2014geotagging}
\bibfield{author}{\bibinfo{person}{Ryan Compton}, \bibinfo{person}{David
  Jurgens}, {and} \bibinfo{person}{David Allen}.}
  \bibinfo{year}{2014}\natexlab{}.
\newblock \showarticletitle{Geotagging One Hundred Million {Twitter} Accounts
  with Total Variation Minimization}. In \bibinfo{booktitle}{\emph{IEEE
  Conference on BigData}}.
\newblock


\bibitem[\protect\citeauthoryear{Conneau, Lample, Ranzato, Denoyer, and
  J{\'e}gou}{Conneau et~al\mbox{.}}{2017}]%
        {conneau2017word}
\bibfield{author}{\bibinfo{person}{Alexis Conneau}, \bibinfo{person}{Guillaume
  Lample}, \bibinfo{person}{Marc'Aurelio Ranzato}, \bibinfo{person}{Ludovic
  Denoyer}, {and} \bibinfo{person}{Herv{\'e} J{\'e}gou}.}
  \bibinfo{year}{2017}\natexlab{}.
\newblock \showarticletitle{Word Translation Without Parallel Data}.
\newblock \bibinfo{journal}{\emph{arXiv preprint arXiv:1710.04087}}
  (\bibinfo{year}{2017}).
\newblock


\bibitem[\protect\citeauthoryear{de~Silva and Compton}{de~Silva and
  Compton}{2014}]%
        {deSilva2014Prediction}
\bibfield{author}{\bibinfo{person}{Brian de Silva} {and} \bibinfo{person}{Ryan
  Compton}.} \bibinfo{year}{2014}\natexlab{}.
\newblock \showarticletitle{Prediction of Foreign Box Office Revenues Based on
  Wikipedia Page Activity}.
\newblock \bibinfo{journal}{\emph{CoRR}}  \bibinfo{volume}{abs/1405.5924}
  (\bibinfo{year}{2014}).
\newblock
\showeprint[arxiv]{1405.5924}
\urldef\tempurl%
\url{http://arxiv.org/abs/1405.5924}
\showURL{%
\tempurl}


\bibitem[\protect\citeauthoryear{DeFrancisco, Palczewski, and
  McGeough}{DeFrancisco et~al\mbox{.}}{2013}]%
        {defrancisco2013gender}
\bibfield{author}{\bibinfo{person}{Victoria~Pruin DeFrancisco},
  \bibinfo{person}{Catherine~Helen Palczewski}, {and}
  \bibinfo{person}{Danielle~D McGeough}.} \bibinfo{year}{2013}\natexlab{}.
\newblock \bibinfo{booktitle}{\emph{Gender in communication: A critical
  introduction}}.
\newblock \bibinfo{publisher}{Sage Publications}.
\newblock


\bibitem[\protect\citeauthoryear{Dehon and Br{\'e}dart}{Dehon and
  Br{\'e}dart}{2001}]%
        {dehon2001other}
\bibfield{author}{\bibinfo{person}{Hedwige Dehon} {and} \bibinfo{person}{Serge
  Br{\'e}dart}.} \bibinfo{year}{2001}\natexlab{}.
\newblock \showarticletitle{An 'other-race' effect in age estimation from
  faces}.
\newblock \bibinfo{journal}{\emph{Perception}} \bibinfo{volume}{30},
  \bibinfo{number}{9} (\bibinfo{year}{2001}), \bibinfo{pages}{1107--1113}.
\newblock


\bibitem[\protect\citeauthoryear{Eckert}{Eckert}{2008}]%
        {eckert2008variation}
\bibfield{author}{\bibinfo{person}{Penelope Eckert}.}
  \bibinfo{year}{2008}\natexlab{}.
\newblock \showarticletitle{Variation and the indexical field}.
\newblock \bibinfo{journal}{\emph{Journal of sociolinguistics}}
  \bibinfo{volume}{12}, \bibinfo{number}{4} (\bibinfo{year}{2008}),
  \bibinfo{pages}{453--476}.
\newblock


\bibitem[\protect\citeauthoryear{Eckert and McConnell-Ginet}{Eckert and
  McConnell-Ginet}{2003}]%
        {eckert2003language}
\bibfield{author}{\bibinfo{person}{Penelope Eckert} {and}
  \bibinfo{person}{Sally McConnell-Ginet}.} \bibinfo{year}{2003}\natexlab{}.
\newblock \bibinfo{booktitle}{\emph{Language and gender}}.
\newblock \bibinfo{publisher}{Cambridge University Press}.
\newblock


\bibitem[\protect\citeauthoryear{{European Commission}}{{European
  Commission}}{[n. d.]}]%
        {eunuts3}
\bibfield{author}{\bibinfo{person}{{European Commission}}.} \bibinfo{year}{[n.
  d.]}\natexlab{}.
\newblock \bibinfo{title}{NUTS---Nomenclature of Territorial Units for
  Statistics}.
\newblock
  \bibinfo{howpublished}{\url{https://ec.europa.eu/eurostat/web/nuts/background}}.
\newblock


\bibitem[\protect\citeauthoryear{Faruqui, Tsvetkov, Neubig, and Dyer}{Faruqui
  et~al\mbox{.}}{2016}]%
        {faruqui2016morphological}
\bibfield{author}{\bibinfo{person}{Manaal Faruqui}, \bibinfo{person}{Yulia
  Tsvetkov}, \bibinfo{person}{Graham Neubig}, {and} \bibinfo{person}{Chris
  Dyer}.} \bibinfo{year}{2016}\natexlab{}.
\newblock \showarticletitle{Morphological inflection generation using character
  sequence to sequence learning}. In \bibinfo{booktitle}{\emph{Proceedings of
  EMNLP}}.
\newblock


\bibitem[\protect\citeauthoryear{Gayo-Avello}{Gayo-Avello}{2012}]%
        {gayo2012wanted}
\bibfield{author}{\bibinfo{person}{Daniel Gayo-Avello}.}
  \bibinfo{year}{2012}\natexlab{}.
\newblock \showarticletitle{" I Wanted to Predict Elections with Twitter and
  all I got was this Lousy Paper"--A Balanced Survey on Election Prediction
  using Twitter Data}.
\newblock \bibinfo{journal}{\emph{arXiv preprint arXiv:1204.6441}}
  (\bibinfo{year}{2012}).
\newblock


\bibitem[\protect\citeauthoryear{Gayo-Avello}{Gayo-Avello}{2013}]%
        {gayo2013meta}
\bibfield{author}{\bibinfo{person}{Daniel Gayo-Avello}.}
  \bibinfo{year}{2013}\natexlab{}.
\newblock \showarticletitle{A meta-analysis of state-of-the-art electoral
  prediction from Twitter data}.
\newblock \bibinfo{journal}{\emph{Social Science Computer Review}}
  \bibinfo{volume}{31}, \bibinfo{number}{6} (\bibinfo{year}{2013}),
  \bibinfo{pages}{649--679}.
\newblock


\bibitem[\protect\citeauthoryear{Generous, Fairchild, Deshpande, Del~Valle, and
  Priedhorsky}{Generous et~al\mbox{.}}{2014}]%
        {Generous2014Global}
\bibfield{author}{\bibinfo{person}{Nicholas Generous},
  \bibinfo{person}{Geoffrey Fairchild}, \bibinfo{person}{Alina Deshpande},
  \bibinfo{person}{Sara~Y. Del~Valle}, {and} \bibinfo{person}{Reid
  Priedhorsky}.} \bibinfo{year}{2014}\natexlab{}.
\newblock \showarticletitle{Global Disease Monitoring and Forecasting with
  Wikipedia}.
\newblock \bibinfo{journal}{\emph{PLOS Computational Biology}}
  \bibinfo{volume}{10}, \bibinfo{number}{11} (\bibinfo{date}{11}
  \bibinfo{year}{2014}), \bibinfo{pages}{1--16}.
\newblock
\urldef\tempurl%
\url{https://doi.org/10.1371/journal.pcbi.1003892}
\showDOI{\tempurl}


\bibitem[\protect\citeauthoryear{Ginsberg, Mohebbi, Patel, Brammer, Smolinski,
  and Brilliant}{Ginsberg et~al\mbox{.}}{2008}]%
        {Ginsberg2008Detecting}
\bibfield{author}{\bibinfo{person}{Jeremy Ginsberg},
  \bibinfo{person}{Matthew~H. Mohebbi}, \bibinfo{person}{Rajan~S. Patel},
  \bibinfo{person}{Lynnette Brammer}, \bibinfo{person}{Mark~S. Smolinski},
  {and} \bibinfo{person}{Larry Brilliant}.} \bibinfo{year}{2008}\natexlab{}.
\newblock \showarticletitle{Detecting influenza epidemics using search engine
  query data}.
\newblock \bibinfo{journal}{\emph{Nature}}  \bibinfo{volume}{457}
  (\bibinfo{date}{nov} \bibinfo{year}{2008}), \bibinfo{pages}{1012}.
\newblock
\urldef\tempurl%
\url{https://doi.org/10.1038/nature07634 10.1038/nature07634}
\showDOI{\tempurl}


\bibitem[\protect\citeauthoryear{Goot, Ljube{\v{s}}i{\'c}, Matroos, Nissim, and
  Plank}{Goot et~al\mbox{.}}{2018}]%
        {goot2018bleaching}
\bibfield{author}{\bibinfo{person}{Rob Goot}, \bibinfo{person}{Nikola
  Ljube{\v{s}}i{\'c}}, \bibinfo{person}{Ian Matroos}, \bibinfo{person}{Malvina
  Nissim}, {and} \bibinfo{person}{Barbara Plank}.}
  \bibinfo{year}{2018}\natexlab{}.
\newblock \showarticletitle{Bleaching Text: Abstract Features for Cross-lingual
  Gender Prediction}. In \bibinfo{booktitle}{\emph{Proceedings of the 56th
  Annual Meeting of the Association for Computational Linguistics}}.
  \bibinfo{pages}{383--389}.
\newblock


\bibitem[\protect\citeauthoryear{Goswami, Sarkar, and Rustagi}{Goswami
  et~al\mbox{.}}{2009}]%
        {goswami2009stylometric}
\bibfield{author}{\bibinfo{person}{Sumit Goswami}, \bibinfo{person}{Sudeshna
  Sarkar}, {and} \bibinfo{person}{Mayur Rustagi}.}
  \bibinfo{year}{2009}\natexlab{}.
\newblock \showarticletitle{Stylometric analysis of bloggers’ age and
  gender}. In \bibinfo{booktitle}{\emph{Proceedings of ICWSM}}.
\newblock


\bibitem[\protect\citeauthoryear{Hale}{Hale}{2014}]%
        {hale2014global}
\bibfield{author}{\bibinfo{person}{Scott~A. Hale}.}
  \bibinfo{year}{2014}\natexlab{}.
\newblock \showarticletitle{Global Connectivity and Multilinguals in the
  Twitter Network}. In \bibinfo{booktitle}{\emph{Proceedings of the SIGCHI
  Conference on Human Factors in Computing Systems}}
  \emph{(\bibinfo{series}{CHI '14})}. \bibinfo{publisher}{ACM},
  \bibinfo{address}{New York, NY, USA}, \bibinfo{pages}{833--842}.
\newblock
\showISBNx{978-1-4503-2473-1}
\urldef\tempurl%
\url{https://doi.org/10.1145/2556288.2557203}
\showDOI{\tempurl}


\bibitem[\protect\citeauthoryear{Hecht and Stephens}{Hecht and
  Stephens}{2014}]%
        {hecht2014urban}
\bibfield{author}{\bibinfo{person}{Brent Hecht} {and} \bibinfo{person}{Monica
  Stephens}.} \bibinfo{year}{2014}\natexlab{}.
\newblock \showarticletitle{A Tale of Cities: Urban Biases in Volunteered
  Geographic Information}. In \bibinfo{booktitle}{\emph{Proceedings of ICWSM}}.
\newblock
\urldef\tempurl%
\url{https://www.aaai.org/ocs/index.php/ICWSM/ICWSM14/paper/view/8114}
\showURL{%
\tempurl}


\bibitem[\protect\citeauthoryear{Hochreiter and Schmidhuber}{Hochreiter and
  Schmidhuber}{1997}]%
        {hochreiter1997long}
\bibfield{author}{\bibinfo{person}{Sepp Hochreiter} {and}
  \bibinfo{person}{J{\"u}rgen Schmidhuber}.} \bibinfo{year}{1997}\natexlab{}.
\newblock \showarticletitle{Long short-term memory}.
\newblock \bibinfo{journal}{\emph{Neural computation}} \bibinfo{volume}{9},
  \bibinfo{number}{8} (\bibinfo{year}{1997}), \bibinfo{pages}{1735--1780}.
\newblock


\bibitem[\protect\citeauthoryear{Holt and Smith}{Holt and Smith}{1979}]%
        {holt1979poststratification}
\bibfield{author}{\bibinfo{person}{D. Holt} {and} \bibinfo{person}{T.~M.~F.
  Smith}.} \bibinfo{year}{1979}\natexlab{}.
\newblock \showarticletitle{Post Stratification}.
\newblock \bibinfo{journal}{\emph{Journal of the Royal Statistical Society.
  Series A (General)}} \bibinfo{volume}{142}, \bibinfo{number}{1}
  (\bibinfo{year}{1979}), \bibinfo{pages}{33--46}.
\newblock
\showISSN{00359238}
\urldef\tempurl%
\url{http://www.jstor.org/stable/2344652}
\showURL{%
\tempurl}


\bibitem[\protect\citeauthoryear{Huang, Liu, Van Der~Maaten, and
  Weinberger}{Huang et~al\mbox{.}}{2017}]%
        {huang2017densely}
\bibfield{author}{\bibinfo{person}{Gao Huang}, \bibinfo{person}{Zhuang Liu},
  \bibinfo{person}{Laurens Van Der~Maaten}, {and} \bibinfo{person}{Kilian~Q
  Weinberger}.} \bibinfo{year}{2017}\natexlab{}.
\newblock \showarticletitle{Densely Connected Convolutional Networks.}. In
  \bibinfo{booktitle}{\emph{CVPR}}, Vol.~\bibinfo{volume}{1}.
  \bibinfo{pages}{3}.
\newblock


\bibitem[\protect\citeauthoryear{Jaech and Ostendorf}{Jaech and
  Ostendorf}{2015}]%
        {jaech2015your}
\bibfield{author}{\bibinfo{person}{Aaron Jaech} {and} \bibinfo{person}{Mari
  Ostendorf}.} \bibinfo{year}{2015}\natexlab{}.
\newblock \showarticletitle{What your username says about you}.
\newblock \bibinfo{journal}{\emph{arXiv preprint arXiv:1507.02045}}
  (\bibinfo{year}{2015}).
\newblock


\bibitem[\protect\citeauthoryear{Jansen}{Jansen}{2010}]%
        {jansen2010use}
\bibfield{author}{\bibinfo{person}{Bernard~J Jansen}.}
  \bibinfo{year}{2010}\natexlab{}.
\newblock \bibinfo{booktitle}{\emph{Use of the internet in higher-income
  households}}.
\newblock \bibinfo{publisher}{Pew Research Center Washington, DC}.
\newblock


\bibitem[\protect\citeauthoryear{Johnson, McMahon, Sch{\"o}ning, and
  Hecht}{Johnson et~al\mbox{.}}{2017}]%
        {johnson2017effect}
\bibfield{author}{\bibinfo{person}{Isaac Johnson}, \bibinfo{person}{Connor
  McMahon}, \bibinfo{person}{Johannes Sch{\"o}ning}, {and}
  \bibinfo{person}{Brent Hecht}.} \bibinfo{year}{2017}\natexlab{}.
\newblock \showarticletitle{The Effect of Population and Structural Biases on
  Social Media-based Algorithms: A Case Study in Geolocation Inference Across
  the Urban-Rural Spectrum}. In \bibinfo{booktitle}{\emph{Proceedings of the
  2017 CHI Conference on Human Factors in Computing Systems}}. ACM,
  \bibinfo{pages}{1167--1178}.
\newblock


\bibitem[\protect\citeauthoryear{Jung, An, Kwak, Salminen, and Jansen}{Jung
  et~al\mbox{.}}{2017}]%
        {jung2017inferring}
\bibfield{author}{\bibinfo{person}{Soon-Gyo Jung}, \bibinfo{person}{Jisun An},
  \bibinfo{person}{Haewoon Kwak}, \bibinfo{person}{Joni Salminen}, {and}
  \bibinfo{person}{Bernard~J Jansen}.} \bibinfo{year}{2017}\natexlab{}.
\newblock \showarticletitle{Inferring Social Media Users' Demographics from
  Profile Pictures: A Face++ Analysis on Twitter Users}. In
  \bibinfo{booktitle}{\emph{Proceedings of The 17th International Conference on
  Electronic Business}}. \bibinfo{pages}{140--145)}.
\newblock


\bibitem[\protect\citeauthoryear{Jungherr}{Jungherr}{2017}]%
        {jungherr2017normalizing}
\bibfield{author}{\bibinfo{person}{Andreas Jungherr}.}
  \bibinfo{year}{2017}\natexlab{}.
\newblock \showarticletitle{Normalizing digital trace data}.
\newblock \bibinfo{journal}{\emph{Digital Discussions: How Big Data Informs
  Political Communication}} (\bibinfo{year}{2017}).
\newblock


\bibitem[\protect\citeauthoryear{Jungherr, J{\"u}rgens, and Schoen}{Jungherr
  et~al\mbox{.}}{2012}]%
        {jungherr2012pirate}
\bibfield{author}{\bibinfo{person}{Andreas Jungherr}, \bibinfo{person}{Pascal
  J{\"u}rgens}, {and} \bibinfo{person}{Harald Schoen}.}
  \bibinfo{year}{2012}\natexlab{}.
\newblock \showarticletitle{Why the pirate party won the german election of
  2009 or the trouble with predictions: {Tumasjan, A., Sprenger, T. O., Sander,
  P. G., \& Welpe, I. M.} ``{P}redicting elections with twitter: What 140
  characters reveal about political sentiment''}.
\newblock \bibinfo{journal}{\emph{Social science computer review}}
  \bibinfo{volume}{30}, \bibinfo{number}{2} (\bibinfo{year}{2012}),
  \bibinfo{pages}{229--234}.
\newblock


\bibitem[\protect\citeauthoryear{Kendall, Tannen, et~al\mbox{.}}{Kendall
  et~al\mbox{.}}{1997}]%
        {kendall1997gender}
\bibfield{author}{\bibinfo{person}{Shari Kendall}, \bibinfo{person}{Deborah
  Tannen}, {et~al\mbox{.}}} \bibinfo{year}{1997}\natexlab{}.
\newblock \showarticletitle{Gender and language in the workplace}.
\newblock \bibinfo{journal}{\emph{Gender and Discourse. London: Sage}}
  (\bibinfo{year}{1997}), \bibinfo{pages}{81--105}.
\newblock


\bibitem[\protect\citeauthoryear{Kim, Jernite, Sontag, and Rush}{Kim
  et~al\mbox{.}}{2016}]%
        {kim2016character}
\bibfield{author}{\bibinfo{person}{Yoon Kim}, \bibinfo{person}{Yacine Jernite},
  \bibinfo{person}{David Sontag}, {and} \bibinfo{person}{Alexander~M Rush}.}
  \bibinfo{year}{2016}\natexlab{}.
\newblock \showarticletitle{Character-Aware Neural Language Models.}. In
  \bibinfo{booktitle}{\emph{AAAI}}. \bibinfo{pages}{2741--2749}.
\newblock


\bibitem[\protect\citeauthoryear{Knowles, Carroll, and Dredze}{Knowles
  et~al\mbox{.}}{2016}]%
        {knowles2016demographer}
\bibfield{author}{\bibinfo{person}{Rebecca Knowles}, \bibinfo{person}{Josh
  Carroll}, {and} \bibinfo{person}{Mark Dredze}.}
  \bibinfo{year}{2016}\natexlab{}.
\newblock \showarticletitle{Demographer: Extremely simple name demographics}.
  In \bibinfo{booktitle}{\emph{Proceedings of the First Workshop on NLP and
  Computational Social Science}}. \bibinfo{pages}{108--113}.
\newblock


\bibitem[\protect\citeauthoryear{Krippendorff}{Krippendorff}{2011}]%
        {krippendorff2011computing}
\bibfield{author}{\bibinfo{person}{Klaus Krippendorff}.}
  \bibinfo{year}{2011}\natexlab{}.
\newblock \showarticletitle{Computing Krippendorff's alpha-reliability}.
\newblock \bibinfo{journal}{\emph{University of Pennsylvania Departmental
  papers (ASC)}} (\bibinfo{year}{2011}).
\newblock


\bibitem[\protect\citeauthoryear{Lakoff and Bucholtz}{Lakoff and
  Bucholtz}{2004}]%
        {lakoff2004language}
\bibfield{author}{\bibinfo{person}{Robin~Tolmach Lakoff} {and}
  \bibinfo{person}{Mary Bucholtz}.} \bibinfo{year}{2004}\natexlab{}.
\newblock \bibinfo{booktitle}{\emph{Language and woman's place: Text and
  commentaries}}. Vol.~\bibinfo{volume}{3}.
\newblock \bibinfo{publisher}{Oxford University Press, USA}.
\newblock


\bibitem[\protect\citeauthoryear{Lamanna, Lenormand, Salas-Olmedo, Romanillos,
  Gon{\c{c}}alves, and Ramasco}{Lamanna et~al\mbox{.}}{2018}]%
        {lamanna2018immigrant}
\bibfield{author}{\bibinfo{person}{Fabio Lamanna}, \bibinfo{person}{Maxime
  Lenormand}, \bibinfo{person}{Mar{\'\i}a~Henar Salas-Olmedo},
  \bibinfo{person}{Gustavo Romanillos}, \bibinfo{person}{Bruno
  Gon{\c{c}}alves}, {and} \bibinfo{person}{Jos{\'e}~J Ramasco}.}
  \bibinfo{year}{2018}\natexlab{}.
\newblock \showarticletitle{Immigrant community integration in world cities}.
\newblock \bibinfo{journal}{\emph{PloS one}} \bibinfo{volume}{13},
  \bibinfo{number}{3} (\bibinfo{year}{2018}), \bibinfo{pages}{e0191612}.
\newblock


\bibitem[\protect\citeauthoryear{Lazer, Kennedy, King, and Vespignani}{Lazer
  et~al\mbox{.}}{2014}]%
        {Lazer2014Parable}
\bibfield{author}{\bibinfo{person}{David Lazer}, \bibinfo{person}{Ryan
  Kennedy}, \bibinfo{person}{Gary King}, {and} \bibinfo{person}{Alessandro
  Vespignani}.} \bibinfo{year}{2014}\natexlab{}.
\newblock \showarticletitle{{The Parable of Google Flu: Traps in Big Data
  Analysis}}.
\newblock \bibinfo{journal}{\emph{Science}} \bibinfo{volume}{343},
  \bibinfo{number}{6167} (\bibinfo{year}{2014}), \bibinfo{pages}{1203--1205}.
\newblock
\showISBNx{0036-8075}
\showISSN{1095-9203}


\bibitem[\protect\citeauthoryear{Little and Rubin}{Little and Rubin}{2002}]%
        {Little2002}
\bibfield{author}{\bibinfo{person}{Roderick Little} {and}
  \bibinfo{person}{Donald Rubin}.} \bibinfo{year}{2002}\natexlab{}.
\newblock \bibinfo{booktitle}{\emph{Statistical analysis with missing data,
  Second edition}}.
\newblock 408 pages.
\newblock
\showISBNx{0471183865}
\showISSN{00324663}


\bibitem[\protect\citeauthoryear{McCandless}{McCandless}{2010}]%
        {cld2}
\bibfield{author}{\bibinfo{person}{Michael McCandless}.}
  \bibinfo{year}{2010}\natexlab{}.
\newblock \bibinfo{title}{Accuracy and performance of {G}oogle's compact
  language detector}.
\newblock
  \bibinfo{howpublished}{\url{http://blog.mikemccandless.com/2011/10/accuracy-and-performance-of-googles.html}}.
\newblock


\bibitem[\protect\citeauthoryear{McCorriston, Jurgens, and Ruths}{McCorriston
  et~al\mbox{.}}{2015}]%
        {mccorriston2015organizations}
\bibfield{author}{\bibinfo{person}{James McCorriston}, \bibinfo{person}{David
  Jurgens}, {and} \bibinfo{person}{Derek Ruths}.}
  \bibinfo{year}{2015}\natexlab{}.
\newblock \showarticletitle{Organizations Are Users Too: Characterizing and
  Detecting the Presence of Organizations on Twitter.}. In
  \bibinfo{booktitle}{\emph{Proceedings of ICWSM}}. \bibinfo{pages}{650--653}.
\newblock


\bibitem[\protect\citeauthoryear{Mesty{\'a}n, Yasseri, and
  Kert{\'e}sz}{Mesty{\'a}n et~al\mbox{.}}{2013}]%
        {Mestyan2013Early}
\bibfield{author}{\bibinfo{person}{M{\'a}rton Mesty{\'a}n},
  \bibinfo{person}{Taha Yasseri}, {and} \bibinfo{person}{János Kert{\'e}sz}.}
  \bibinfo{year}{2013}\natexlab{}.
\newblock \showarticletitle{Early Prediction of Movie Box Office Success Based
  on Wikipedia Activity Big Data}.
\newblock \bibinfo{journal}{\emph{PLOS ONE}} \bibinfo{volume}{8},
  \bibinfo{number}{8} (\bibinfo{date}{08} \bibinfo{year}{2013}),
  \bibinfo{pages}{1--8}.
\newblock
\urldef\tempurl%
\url{https://doi.org/10.1371/journal.pone.0071226}
\showDOI{\tempurl}


\bibitem[\protect\citeauthoryear{Mislove, Lehmann, Ahn, Onnela, and
  Rosenquist}{Mislove et~al\mbox{.}}{2011}]%
        {mislove2011understanding}
\bibfield{author}{\bibinfo{person}{Alan Mislove}, \bibinfo{person}{Sune
  Lehmann}, \bibinfo{person}{Yong-Yeol Ahn}, \bibinfo{person}{Jukka-Pekka
  Onnela}, {and} \bibinfo{person}{J~Niels Rosenquist}.}
  \bibinfo{year}{2011}\natexlab{}.
\newblock \showarticletitle{Understanding the Demographics of Twitter Users.}
\newblock \bibinfo{journal}{\emph{Proceedings of ICWSM}} \bibinfo{volume}{11},
  \bibinfo{number}{5th} (\bibinfo{year}{2011}), \bibinfo{pages}{25}.
\newblock


\bibitem[\protect\citeauthoryear{Nair and Hinton}{Nair and Hinton}{2010}]%
        {nair2010rectified}
\bibfield{author}{\bibinfo{person}{Vinod Nair} {and}
  \bibinfo{person}{Geoffrey~E Hinton}.} \bibinfo{year}{2010}\natexlab{}.
\newblock \showarticletitle{Rectified linear units improve restricted boltzmann
  machines}. In \bibinfo{booktitle}{\emph{Proceedings of the 27th international
  conference on machine learning (ICML-10)}}. \bibinfo{pages}{807--814}.
\newblock


\bibitem[\protect\citeauthoryear{Nguyen, Gravel, Trieschnigg, and Meder}{Nguyen
  et~al\mbox{.}}{2013}]%
        {nguyen2013old}
\bibfield{author}{\bibinfo{person}{Dong Nguyen}, \bibinfo{person}{Rilana
  Gravel}, \bibinfo{person}{Dolf Trieschnigg}, {and} \bibinfo{person}{Theo
  Meder}.} \bibinfo{year}{2013}\natexlab{}.
\newblock \showarticletitle{``How Old Do You Think I Am?'' A Study of Language
  and Age in Twitter.}. In \bibinfo{booktitle}{\emph{Proceedings of ICWSM}}.
\newblock


\bibitem[\protect\citeauthoryear{Nguyen, Smith, and Ros{\'e}}{Nguyen
  et~al\mbox{.}}{2011}]%
        {nguyen2011author}
\bibfield{author}{\bibinfo{person}{Dong Nguyen}, \bibinfo{person}{Noah~A
  Smith}, {and} \bibinfo{person}{Carolyn~P Ros{\'e}}.}
  \bibinfo{year}{2011}\natexlab{}.
\newblock \showarticletitle{Author age prediction from text using linear
  regression}. In \bibinfo{booktitle}{\emph{Proc. of the Workshop on Language
  Technology for Cultural Heritage, Social Sciences, and Humanities}}.
  Association for Computational Linguistics, \bibinfo{pages}{115--123}.
\newblock


\bibitem[\protect\citeauthoryear{Park, Gelman, and Bafumi}{Park
  et~al\mbox{.}}{2004}]%
        {park2004bayesian}
\bibfield{author}{\bibinfo{person}{David~K Park}, \bibinfo{person}{Andrew
  Gelman}, {and} \bibinfo{person}{Joseph Bafumi}.}
  \bibinfo{year}{2004}\natexlab{}.
\newblock \showarticletitle{Bayesian multilevel estimation with
  poststratification: state-level estimates from national polls}.
\newblock \bibinfo{journal}{\emph{Political Analysis}} \bibinfo{volume}{12},
  \bibinfo{number}{4} (\bibinfo{year}{2004}), \bibinfo{pages}{375--385}.
\newblock


\bibitem[\protect\citeauthoryear{Paszke, Gross, Chintala, Chanan, Yang, DeVito,
  Lin, Desmaison, Antiga, and Lerer}{Paszke et~al\mbox{.}}{2017}]%
        {paszke2017automatic}
\bibfield{author}{\bibinfo{person}{Adam Paszke}, \bibinfo{person}{Sam Gross},
  \bibinfo{person}{Soumith Chintala}, \bibinfo{person}{Gregory Chanan},
  \bibinfo{person}{Edward Yang}, \bibinfo{person}{Zachary DeVito},
  \bibinfo{person}{Zeming Lin}, \bibinfo{person}{Alban Desmaison},
  \bibinfo{person}{Luca Antiga}, {and} \bibinfo{person}{Adam Lerer}.}
  \bibinfo{year}{2017}\natexlab{}.
\newblock \showarticletitle{Automatic differentiation in PyTorch}. In
  \bibinfo{booktitle}{\emph{NIPS-W}}.
\newblock


\bibitem[\protect\citeauthoryear{Rangel, Rosso, Montes-y G{\'o}mez, Potthast,
  and Stein}{Rangel et~al\mbox{.}}{2018}]%
        {rangel2018overview}
\bibfield{author}{\bibinfo{person}{Francisco Rangel}, \bibinfo{person}{Paolo
  Rosso}, \bibinfo{person}{Manuel Montes-y G{\'o}mez}, \bibinfo{person}{Martin
  Potthast}, {and} \bibinfo{person}{Benno Stein}.}
  \bibinfo{year}{2018}\natexlab{}.
\newblock \showarticletitle{Overview of the 6th author profiling task at PAN
  2018: Multimodal gender identification in Twitter}.
\newblock \bibinfo{journal}{\emph{Working Notes Papers of the CLEF}}
  (\bibinfo{year}{2018}).
\newblock


\bibitem[\protect\citeauthoryear{Rao, Yarowsky, Shreevats, and Gupta}{Rao
  et~al\mbox{.}}{2010}]%
        {rao2010classifying}
\bibfield{author}{\bibinfo{person}{Delip Rao}, \bibinfo{person}{David
  Yarowsky}, \bibinfo{person}{Abhishek Shreevats}, {and}
  \bibinfo{person}{Manaswi Gupta}.} \bibinfo{year}{2010}\natexlab{}.
\newblock \showarticletitle{Classifying latent user attributes in twitter}. In
  \bibinfo{booktitle}{\emph{Proc. of the 2nd International Workshop on Search
  and Mining User-generated Contents}}. ACM, \bibinfo{pages}{37--44}.
\newblock


\bibitem[\protect\citeauthoryear{Reddi, Kale, and Kumar}{Reddi
  et~al\mbox{.}}{2018}]%
        {reddi2018convergence}
\bibfield{author}{\bibinfo{person}{Sashank~J Reddi}, \bibinfo{person}{Satyen
  Kale}, {and} \bibinfo{person}{Sanjiv Kumar}.}
  \bibinfo{year}{2018}\natexlab{}.
\newblock \showarticletitle{On the convergence of adam and beyond}.
\newblock  (\bibinfo{year}{2018}).
\newblock


\bibitem[\protect\citeauthoryear{Rosenthal and McKeown}{Rosenthal and
  McKeown}{2011}]%
        {rosenthal2011age}
\bibfield{author}{\bibinfo{person}{Sara Rosenthal} {and}
  \bibinfo{person}{Kathleen McKeown}.} \bibinfo{year}{2011}\natexlab{}.
\newblock \showarticletitle{Age prediction in blogs: A study of style, content,
  and online behavior in pre-and post-social media generations}. In
  \bibinfo{booktitle}{\emph{Proc. ACL}}. Association for Computational
  Linguistics, \bibinfo{pages}{763--772}.
\newblock


\bibitem[\protect\citeauthoryear{Rothe, Timofte, and Gool}{Rothe
  et~al\mbox{.}}{2016}]%
        {rothe2016deep}
\bibfield{author}{\bibinfo{person}{Rasmus Rothe}, \bibinfo{person}{Radu
  Timofte}, {and} \bibinfo{person}{Luc~Van Gool}.}
  \bibinfo{year}{2016}\natexlab{}.
\newblock \showarticletitle{Deep expectation of real and apparent age from a
  single image without facial landmarks}.
\newblock \bibinfo{journal}{\emph{International Journal of Computer Vision
  (IJCV)}} (\bibinfo{date}{July} \bibinfo{year}{2016}).
\newblock


\bibitem[\protect\citeauthoryear{Ruths and Pfeffer}{Ruths and Pfeffer}{2014}]%
        {ruths2014social}
\bibfield{author}{\bibinfo{person}{Derek Ruths} {and}
  \bibinfo{person}{J{\"u}rgen Pfeffer}.} \bibinfo{year}{2014}\natexlab{}.
\newblock \showarticletitle{Social media for large studies of behavior}.
\newblock \bibinfo{journal}{\emph{Science}} \bibinfo{volume}{346},
  \bibinfo{number}{6213} (\bibinfo{year}{2014}), \bibinfo{pages}{1063--1064}.
\newblock


\bibitem[\protect\citeauthoryear{Sap, Park, Eichstaedt, Kern, Stillwell,
  Kosinski, Ungar, and Schwartz}{Sap et~al\mbox{.}}{2014}]%
        {sap2014developing}
\bibfield{author}{\bibinfo{person}{Maarten Sap}, \bibinfo{person}{Gregory
  Park}, \bibinfo{person}{Johannes Eichstaedt}, \bibinfo{person}{Margaret
  Kern}, \bibinfo{person}{David Stillwell}, \bibinfo{person}{Michal Kosinski},
  \bibinfo{person}{Lyle Ungar}, {and} \bibinfo{person}{H.~Andrew Schwartz}.}
  \bibinfo{year}{2014}\natexlab{}.
\newblock \showarticletitle{Developing Age and Gender Predictive Lexica over
  Social Media}. In \bibinfo{booktitle}{\emph{Proc. EMNLP}}. Association for
  Computational Linguistics, \bibinfo{pages}{1146--1151}.
\newblock


\bibitem[\protect\citeauthoryear{S{\"{a}}rndal and
  Lundstr{\"{o}}m}{S{\"{a}}rndal and Lundstr{\"{o}}m}{2005}]%
        {Sarndal2005}
\bibfield{author}{\bibinfo{person}{Carl-Erik S{\"{a}}rndal} {and}
  \bibinfo{person}{Sixten Lundstr{\"{o}}m}.} \bibinfo{year}{2005}\natexlab{}.
\newblock \bibinfo{booktitle}{\emph{{Estimation in Surveys with Nonresponse}}}.
\newblock \bibinfo{publisher}{John Wiley {\&} Sons, Ltd},
  \bibinfo{address}{Chichester, UK}. 1--199 pages.
\newblock
\showISBNx{9780470011355}
\showISSN{0040-1706}
\urldef\tempurl%
\url{https://doi.org/10.1002/0470011351}
\showDOI{\tempurl}


\bibitem[\protect\citeauthoryear{Schler, Koppel, Argamon, and
  Pennebaker}{Schler et~al\mbox{.}}{2006}]%
        {schler2006effects}
\bibfield{author}{\bibinfo{person}{Jonathan Schler}, \bibinfo{person}{Moshe
  Koppel}, \bibinfo{person}{Shlomo Argamon}, {and} \bibinfo{person}{James~W
  Pennebaker}.} \bibinfo{year}{2006}\natexlab{}.
\newblock \showarticletitle{Effects of age and gender on blogging.}. In
  \bibinfo{booktitle}{\emph{AAAI spring symposium: Computational approaches to
  analyzing weblogs}}, Vol.~\bibinfo{volume}{6}. \bibinfo{pages}{199--205}.
\newblock


\bibitem[\protect\citeauthoryear{Sloan}{Sloan}{2017}]%
        {Sloan2017Who}
\bibfield{author}{\bibinfo{person}{Luke Sloan}.}
  \bibinfo{year}{2017}\natexlab{}.
\newblock \showarticletitle{Who Tweets in the United Kingdom? Profiling the
  Twitter Population Using the British Social Attitudes Survey 2015}.
\newblock \bibinfo{journal}{\emph{Social Media + Society}} \bibinfo{volume}{3},
  \bibinfo{number}{1} (\bibinfo{year}{2017}),
  \bibinfo{pages}{2056305117698981}.
\newblock


\bibitem[\protect\citeauthoryear{Tannen}{Tannen}{1991}]%
        {tannen1991you}
\bibfield{author}{\bibinfo{person}{Deborah Tannen}.}
  \bibinfo{year}{1991}\natexlab{}.
\newblock \bibinfo{booktitle}{\emph{You just don't understand: Women and men in
  conversation}}.
\newblock \bibinfo{publisher}{Virago London}.
\newblock


\bibitem[\protect\citeauthoryear{Tannen}{Tannen}{1993}]%
        {tannen1993gender}
\bibfield{author}{\bibinfo{person}{Deborah Tannen}.}
  \bibinfo{year}{1993}\natexlab{}.
\newblock \bibinfo{booktitle}{\emph{Gender and conversational interaction}}.
\newblock \bibinfo{publisher}{Oxford University Press}.
\newblock


\bibitem[\protect\citeauthoryear{UK Office for National Statistics}{UK Office
  for National Statistics}{2015}]%
        {harmonised2015ukstats}
UK Office for National Statistics \bibinfo{year}{2015}\natexlab{}.
\newblock \bibinfo{booktitle}{\emph{Harmonised Concepts and Questions for
  Social Data Sources - Primary Principles}}.
\newblock UK Office for National Statistics.
\newblock
\urldef\tempurl%
\url{http://www.ons.gov.uk/ons/guide-method/harmonisation/primary-set-of-harmonised-concepts-and-questions/demographic-information--household-composition-and-relationships.pdf}
\showURL{%
\tempurl}


\bibitem[\protect\citeauthoryear{United States Department of Education}{United
  States Department of Education}{2009}]%
        {implementation2009usde}
United States Department of Education \bibinfo{year}{2009}\natexlab{}.
\newblock \bibinfo{booktitle}{\emph{Implementation guidelines: Measures and
  methods for the national reporting system for adult education}}.
\newblock United States Department of Education.
\newblock
\urldef\tempurl%
\url{http://www.air.org/sites/default/files/downloads/report/
  ImplementationGuidelines_0.pdf}
\showURL{%
\tempurl}


\bibitem[\protect\citeauthoryear{Van~Rijsbergen, Jaworska, Rousselet, and
  Schyns}{Van~Rijsbergen et~al\mbox{.}}{2014}]%
        {van2014age}
\bibfield{author}{\bibinfo{person}{Nicola Van~Rijsbergen},
  \bibinfo{person}{Katarzyna Jaworska}, \bibinfo{person}{Guillaume~A
  Rousselet}, {and} \bibinfo{person}{Philippe~G Schyns}.}
  \bibinfo{year}{2014}\natexlab{}.
\newblock \showarticletitle{With age comes representational wisdom in social
  signals}.
\newblock \bibinfo{journal}{\emph{Current Biology}} \bibinfo{volume}{24},
  \bibinfo{number}{23} (\bibinfo{year}{2014}), \bibinfo{pages}{2792--2796}.
\newblock


\bibitem[\protect\citeauthoryear{Wang, Rothschild, Goel, and Gelman}{Wang
  et~al\mbox{.}}{2015}]%
        {wang2015forecasting}
\bibfield{author}{\bibinfo{person}{Wei Wang}, \bibinfo{person}{David
  Rothschild}, \bibinfo{person}{Sharad Goel}, {and} \bibinfo{person}{Andrew
  Gelman}.} \bibinfo{year}{2015}\natexlab{}.
\newblock \showarticletitle{Forecasting elections with non-representative
  polls}.
\newblock \bibinfo{journal}{\emph{International Journal of Forecasting}}
  \bibinfo{volume}{31}, \bibinfo{number}{3} (\bibinfo{year}{2015}),
  \bibinfo{pages}{980--991}.
\newblock


\bibitem[\protect\citeauthoryear{Wang and Jurgens}{Wang and Jurgens}{2018}]%
        {wang2018its}
\bibfield{author}{\bibinfo{person}{Zijian Wang} {and} \bibinfo{person}{David
  Jurgens}.} \bibinfo{year}{2018}\natexlab{}.
\newblock \showarticletitle{It's going to be okay: Measuring Access to Support
  in Online Communities}. In \bibinfo{booktitle}{\emph{Proceedings of the 2018
  Conference on Empirical Methods in Natural Language Processing}}.
  \bibinfo{pages}{33--45}.
\newblock


\bibitem[\protect\citeauthoryear{Wood-Doughty, Mahajan, and
  Dredze}{Wood-Doughty et~al\mbox{.}}{2018}]%
        {wood2018johns}
\bibfield{author}{\bibinfo{person}{Zach Wood-Doughty},
  \bibinfo{person}{Praateek Mahajan}, {and} \bibinfo{person}{Mark Dredze}.}
  \bibinfo{year}{2018}\natexlab{}.
\newblock \showarticletitle{Johns Hopkins or johnny-hopkins: Classifying
  Individuals versus Organizations on Twitter}. In
  \bibinfo{booktitle}{\emph{Proceedings of the Second Workshop on Computational
  Modeling of People's Opinions, Personality, and Emotions in Social Media}}.
  \bibinfo{pages}{56--61}.
\newblock


\bibitem[\protect\citeauthoryear{Wu and Wang}{Wu and Wang}{2007}]%
        {wu2007pivot}
\bibfield{author}{\bibinfo{person}{Hua Wu} {and} \bibinfo{person}{Haifeng
  Wang}.} \bibinfo{year}{2007}\natexlab{}.
\newblock \showarticletitle{Pivot language approach for phrase-based
  statistical machine translation}.
\newblock \bibinfo{journal}{\emph{Machine Translation}} \bibinfo{volume}{21},
  \bibinfo{number}{3} (\bibinfo{year}{2007}), \bibinfo{pages}{165--181}.
\newblock


\bibitem[\protect\citeauthoryear{Zagheni, Weber, and Gummadi}{Zagheni
  et~al\mbox{.}}{2017}]%
        {Zagheni2017Leveraging}
\bibfield{author}{\bibinfo{person}{Emilio Zagheni}, \bibinfo{person}{Ingmar
  Weber}, {and} \bibinfo{person}{Krishna Gummadi}.}
  \bibinfo{year}{2017}\natexlab{}.
\newblock \showarticletitle{{Leveraging Facebook's Advertising Platform to
  Monitor Stocks of Migrants}}.
\newblock \bibinfo{journal}{\emph{Population and Development Review}}
  \bibinfo{volume}{43}, \bibinfo{number}{4} (\bibinfo{year}{2017}),
  \bibinfo{pages}{721--734}.
\newblock
\showISSN{17284457}


\bibitem[\protect\citeauthoryear{Zhang, Hu, Zhang, and Liu}{Zhang
  et~al\mbox{.}}{2016}]%
        {zhang2016your}
\bibfield{author}{\bibinfo{person}{Jinxue Zhang}, \bibinfo{person}{Xia Hu},
  \bibinfo{person}{Yanchao Zhang}, {and} \bibinfo{person}{Huan Liu}.}
  \bibinfo{year}{2016}\natexlab{}.
\newblock \showarticletitle{Your Age Is No Secret: Inferring Microbloggers'
  Ages via Content and Interaction Analysis.}. In
  \bibinfo{booktitle}{\emph{Proceedings of ICWSM}}. \bibinfo{pages}{476--485}.
\newblock


\end{thebibliography}

\end{document}